\documentclass[prx,aps,superscriptaddress,twocolumn,notitlepage,floatfix,10pt]{revtex4-2}
\usepackage{amsmath,mathtools,amsthm,amssymb,pifont}
\usepackage[utf8]{inputenc}
\usepackage[american]{babel}
\usepackage{graphicx,xcolor,bbold,titlesec}
\usepackage{MnSymbol}

\usepackage[colorlinks,
bookmarksopen,
bookmarksnumbered,
citecolor=teal,
linkcolor=teal,
urlcolor=teal]{hyperref}
\usepackage{braket}
\usepackage{enumitem}

\usepackage{ifthen}

\newcommand{\Andrea}[2][]{%
  \textcolor{orange}{#2}%
  \ifthenelse{\equal{#1}{}}{}{%
    \textcolor{orange}{\textbf{[Andrea: #1]}}%
  }%
}
\usepackage{orcidlink}
\usepackage{tikz,ifthen}
\usepackage{bbold}
\usepackage{tikz-network}
\usetikzlibrary{patterns,decorations.pathreplacing,calligraphy}
\usetikzlibrary{shapes,arrows.meta,decorations.pathmorphing}

\newcommand{\kket}[1]{|#1\rrangle}
\newcommand{\bbra}[1]{\llangle #1|}
\newcommand{\bbrakket}[2]{\llangle #1|#2\rrangle}

\newcommand{\titleinfo}{Anticoncentration in Clifford Circuits and Beyond: \\
From Random Tensor Networks to Pseudo-Magic States}

\begin{document}
\title{\titleinfo} 

\author{Beatrice Magni~\orcidlink{0009-0009-8577-0525}}
\email{bmagni@uni-koeln.de}
\affiliation{Institut für Theoretische Physik, Universität zu Köln, Zülpicher Strasse 77, 50937 Köln, Germany}

\author{Alexios Christopoulos~\orcidlink{0000-0003-4857-5922}}
\affiliation{Jožef Stefan Institute, 1000 Ljubljana, Slovenia}

\author{Andrea De Luca~\orcidlink{0000-0003-0272-5083}}
\affiliation{Laboratoire de Physique Théorique et Modélisation, CNRS UMR 8089, CY Cergy Paris Université, F-95302 Cergy-Pontoise, France}

\author{Xhek Turkeshi~\orcidlink{0000-0003-1093-3771}}
\email{xturkesh@uni-koeln.de}
\affiliation{Institut für Theoretische Physik, Universität zu Köln, Zülpicher Strasse 77, 50937 Köln, Germany}

\begin{abstract}
Anticoncentration describes how an ensemble of quantum states spreads over the allowed Hilbert space, leading to statistically uniform output probability distributions. 
In this work, 
we investigate the anticoncentration of random Clifford circuits toward the overlap distribution of random stabilizer states. 
Using exact analytical techniques and extensive numerical simulations based on Clifford replica tensor networks, we demonstrate that random Clifford circuits fully anticoncentrate in logarithmic circuit depth, namely higher-order moments of the overlap distribution converge to those of random stabilizer states. 
Moreover, we investigate the effect of introducing a controlled number of non-Clifford (magic) resources into Clifford circuits. We show that inserting a polylogarithmic in qudit number of $T$-states is sufficient to drive the overlap distribution toward the Porter-Thomas statistics, effectively recovering full quantum randomness. 
In short, this fact presents doped tensor networks and shallow Clifford circuits as pseudo-magic quantum states. 
Our results clarify the interplay between Clifford dynamics, magic-state injection, and quantum complexity, with implications for quantum circuit sampling, many-body quantum physics, and the benchmarking of quantum computational advantage. 
\end{abstract}

\maketitle

\section{Introduction}
Quantum circuits provide a natural framework for describing digital quantum dynamics, serving both as the backbone for quantum computation and as a versatile toolbox to investigate complexity in quantum many-body systems -- from quantum chaos to entanglement spreading~\cite{preskill2018quantum,fisher2023random}. 
A key concept in this context is \textit{anticoncentration}, which measures how spread is the overlap distribution of an ensemble of quantum states in the computational basis~\cite{Bouland2018on,boixo2018characterizing,dalzell2022random}. 
Originally, anticoncentration was introduced as the property of a circuit ensemble to produce sufficiently spread states in Hilbert space. Already an approximate 2-design ensures that 
the overlap with 
any state cannot be too small~\cite{Hangleiter2018anticoncentration}. 
Here, we loosely employ the term anticoncentration to describe \textit{the overlap distribution} for a family of circuit ensembles that increasingly expand over the Hilbert space~\cite{lami2025anticoncentration,turkeshi2024hilbert,christopoulos2024universal}. 
Related to Hilbert space delocalization~\cite{luitz2014universal,mace2019multifractal,sierant2022universal,claeys2025fockspace}, anticoncentration is intimately connected to concepts of quantum randomness and plays a critical role in practical setups, from shallow shadows in randomized measurements on quantum chips~\cite{Huang2020predicting,bertoni2024shallow,arienzo2023closedform,ippoliti2023operator,cioli2024approximate} to benchmarking computational quantum advantage in quantum circuit sampling~\cite{arute2019quantum,hangleiter2023computational,morvan2024phase,ware2023sharp}. 
For chaotic unitary circuits~\cite{nahum2017quantum,nahum2018operator,Keyserlingk2018operator}, anticoncentration has been shown to occur rapidly, with the overlap, or output, distribution approaching the Porter-Thomas one~\cite{porter1956fluctuations} in logarithmic circuit depth~\cite{schuster2025randomunitariesextremelylow,fefferman2024anticoncentrationunitaryhaarmeasure}. 
However, when structural limitations are imposed on quantum circuits, this behavior can be significantly altered. 
A well-known example is Hamiltonian evolution, where energy conservation prevents full anticoncentration~\cite{tirrito2024anticoncentrationmagicspreadingergodic,collura2025quantummagicfermionicgaussian}.
This motivates the question of how anticoncentration occurs in structured quantum many-body dynamics.

In this work, we present a comprehensive analysis of anticoncentration for \textit{random Clifford circuits} applied to stabilizer states, a crucial class of many-body systems with significant applications in quantum error correction and fault-tolerant quantum computing~\cite{gottesman1998theory,gottesman1997stabilizer,nielsen2010quantum}. 
Notably, stabilizer states are efficiently simulatable on classical computers~\cite{aaronson2004improved,Gidney2021stimfaststabilizer,Bravyi2019simulationofquantum} and can be learned with polynomial resources~\cite{rocchetto2018stabiliserstatesefficientlypaclearnable,montanaro2017learningstabilizerstatesbell}, thus they cannot provide computational quantum advantage. 
Interestingly, Clifford unitaries form a 2-design for generic qudit systems -- and a 3-design for qubit systems, implying that the first few statistical moments of their overlap distribution match those of Haar-random unitary gates~\cite{zhu2016cliffordgroupfailsgracefully,gross2021schurweyl}. 
As a result, the intricate structure of the overlap distribution, which underlies the efficient sampling of Clifford circuits, can only be captured through its higher-order moments.

We address and solve the following problems:
\begin{enumerate}[leftmargin=10pt, itemsep=0pt, parsep=0pt, label=\roman*,topsep=3pt]
    \item What is the overlap distribution for random stabilizer states?
    \item What bond dimension is required for a Clifford random tensor network to approximate this distribution? On what timescale is this distribution reached by \emph{local} random Clifford circuits? 
    \item How many non-Clifford (magic) resources~\cite{bravyi2005universal,liu2022manybody,chitambar2019quantum} are needed to drive the overlap statistics toward the unitary Porter-Thomas distribution, characteristic of Hilbert space random states?
\end{enumerate}
To answer these questions, we employ a combination of exact analytical methods and extensive numerical simulations based on Clifford replica tensor networks. 
After determining the explicit form of the overlap distribution for random stabilizer states, we demonstrate that Clifford random matrix product states saturate this distribution with a bond dimension that is polynomial in the system size. 
From these results, inspired by recent works on random unitary circuits~\cite{schuster2025randomunitariesextremelylow}, we show that ``glued'' Clifford circuits anticoncentrate to random stabilizer states in logarithmic depth. 
We corroborate these results with numerical simulations on Clifford brickwork circuits up to $N=512$ qudits. 
Finally, we explore how many magic state resources we must inject into our system to drive the overlap distribution toward the Porter-Thomas statistics. 
Surprisingly, we find that in the thermodynamic limit, a $O(\log(N))$ circuit depth with a logarithmic in qudit number $N$ of magic states suffices. 
At finite sizes, a polylogarithmic number is required to achieve indistinguishability from the Porter-Thomas distribution at any polynomial accuracy, connecting to the concept of pseudo-magic states~\cite{gu2024pseudomagic}. 
These findings provide insights into the relationship between Clifford circuits and quantum complexity, with implications for quantum sampling problems and the benchmarking of computational advantage in near-term quantum computers.

The remainder of the paper is structured as follows. In Sec.~\ref{sec:methods}, we review the mathematical background necessary for our results, covering stabilizer states, Clifford operations, and the Clifford Weingarten calculus. In Sec.~\ref{sec:results}, we present our main findings, including the Clifford analog of the Porter-Thomas distribution and the anticoncentration properties of random tensor network states and circuits. Additionally, we explore the role of non-Clifford resources in enhancing anticoncentration toward the Haar unitary ensemble. Finally, in Sec.~\ref{sec:disc}, we summarize our conclusions and discuss the implication of our work within and beyond quantum information science, and potential directions for future research. 
For self-consistency, App.~\ref{app:A} provides a brief review of Haar-random unitary circuits and their anticoncentration properties.

\section{Methods}
\label{sec:methods}
\subsection{Preliminaries}
We consider a system of $N$ qudits with local Hilbert space $\mathcal{H}_d$ of dimension $d$ a prime number, and denote $\mathcal{H}_{N,d}\equiv \mathcal{H}_d^{\otimes N}$ the total Hilbert space.
The set of Pauli strings 
\begin{equation}
    \mathcal{P}_{N,d}=\{ X_1^{a_1} Z_1^{b_1} X_2^{a_2} Z_2^{b_2} \cdots X_N^{a_N} Z_N^{b_N} \;|\; a_i,b_j \in \mathbb{Z}_d \}\;,
    \label{eq:pauligroup}
\end{equation}
consists of all the $d^{2N}$ operators defined in terms of the generalized Pauli matrices $X = \sum_{m=0}^{d-1} |m\rangle \langle m\oplus 1|$ and $Z = \sum_{m=0}^{d-1} \omega^m |m\rangle \langle m|$ with $\omega \equiv e^{-2\pi i/d}$ and ${a\oplus b\equiv (a+b)\,\mathrm{mod}\,d}$ the addition in the Galois field $\mathbb{Z}_d$. 
The Clifford group on $N$ qudits is the subgroup of the unitary operations $\mathcal{C}_{N,d}\subset \mathcal{U}(d^N)$ that maps a Pauli string $P$ to another Pauli string $\tilde{P}$ up to a phase~\cite{gross2021schurweyl}, namely
\begin{equation}
    C P C^\dagger = \omega^m \tilde{P}, \qquad m \in \mathbb{Z}_d, \qquad P,\tilde P \in \mathcal{P}_{N,d} \;.
\end{equation}
It is generated by the action of the Hadamard $H$, control addition CADD, and phase $S$ gates, respectively
\begin{equation}
\begin{split}
    H&=\frac{1}{\sqrt{d}}\sum_{m,n=0}^{d-1} \omega^{mn} |m\rangle \langle n|\;,\\
    \mathrm{CADD}&=\sum_{m,n=0}^{d-1} |m, m\oplus n\rangle \langle m, n|\;,
\end{split}
\end{equation}
while $S=\sum_{m=0}^{d-1} \omega^{m(m-1) (d+1)/2} |m\rangle \langle m|$ for $d>2$ and \(S = \sqrt{Z}\) for $d=2$. 

Stabilizer states over $N$ qudits are defined by applying a Clifford transformation to $|\pmb{0}\rangle\equiv |0\rangle^{\otimes N}$, namely
\begin{equation}
    \mathrm{Stab}_{N,d}=\{ C\,|\pmb{0}\rangle\;|\; C\in\mathcal{C}_{N,d}\}\;.
\end{equation}
Equivalently, stabilizer states are uniquely characterized by $N$ independent signed Pauli strings, 
$\tilde{P}_j|\Psi\rangle=|\Psi\rangle$ for $j=1,\dots,N$, where $\tilde{P}_j=\omega^{\phi^{(j)}} P_j$ with $\phi^{(j)}\in \mathbb{Z}_d$ and $P_j\in\mathcal{P}_{N,d}$. 
In fact, since $Z_j|\pmb{0}\rangle = |\pmb{0}\rangle$ at each site, it follows that for any Clifford $C$ the stabilizer state $|\Psi\rangle=C|\pmb{0}\rangle$ is stabilized by $\tilde{P}_j = \omega^{n_j} C Z_j C^\dagger$. 
Thus, the knowledge of the generators $\tilde{P}_j$ is equivalent to the complete knowledge of the state, explicitly $|\Psi\rangle\langle \Psi| = \prod_{j=1}^N [\sum_{a=0}^{d-1} \tilde{P}_j^a/d]$. 
As a result, the state of the system is uniquely fixed by a $(2N+1)\times N$ matrix, the so-called stabilizer tableau, $\mathsf{S}=(\phi^{(j)} | \mathbf{a}^{(j)}| \mathbf{b}^{(j)})_{j=1,\dots,N}$, whose rows contain, for each stabilizing string $\tilde P_j$, the parameters $\phi^{(j)}\in \mathbb{Z}_d$ and $\mathbf{a},\, \mathbf{b} \in \mathbb{Z}_d^{N}$ fixing unambiguously the Pauli string~\eqref{eq:pauligroup}. For a given stabilizer state, for later convenience, we define the $N\times N$ submatrices $\mathsf{X}=(a^{(i)}_j)_{i,j=1,\dots,N}$ and $\mathsf{Z}=(b^{(i)}_j)_{i,j=1,\dots,N}$. 
The preceding discussion provides an indication of why Clifford operations are efficiently simulatable in $\mathrm{poly}(N)$ resources~\cite{aaronson2004improved,Bravyi2019simulationofquantum,Gidney2021stimfaststabilizer}, as they map stabilizer $\mathsf{S}$ tableau describing $|\Psi\rangle$ into stabilizer tableau $\mathsf{S}\mapsto \mathsf{S}'$ describing $|\Psi'\rangle=C|\Psi\rangle$.

We conclude this section by setting up notation and conventions we will employ in the rest of this work. With a slight abuse of notation, we will pass from integers ${\pmb{x}=0,\ldots,d^N-1}$ to strings ${\pmb{x}\in \mathbb{Z}_d^N}$ implicitly, assuming the digit to integer convention is inferred. 
Additionally, we will employ the double space representation fixed by $A \rightarrow \kket{A}$ with $\bbrakket{A}{B} =\mathrm{tr}(A^\dagger B)$, and $|U A U^\dagger \rrangle= (U\otimes U^*)\kket{A}$~\cite{Jamiokowski1972}. 
For simplicity, we move between these representations as required, inferring the specific notation from the context.
Lastly, we emphasize that several results tie to \emph{q-analog} arithmetic, for which we reference Ref.~\cite{Kac2002} for a comprehensive discussion.

\subsection{Clifford Weingarten calculus}\label{subsec:methodsb}
Consider a Clifford circuit expressed as the product of Clifford gates acting on contiguous qudits. Explicitly, the circuit is organized into layers identified by the time, or depth, $s$. 
We can express the whole circuit by
\begin{equation}
    \mathbf{C}_t = \prod_{s=0}^t\left(\prod_{\lambda\in \Lambda_s}C_{\lambda}\right)\;,\label{eq:circuit}
\end{equation}
where $\Lambda_s$ fix the action of gates for the $s$-th layer and the $C_{\lambda}$ are Clifford gates acting on the set $\lambda$ of contiguous qudits \footnote{Explicitly $C_\lambda \equiv C_{\lambda_1, \ldots, \lambda_n} = \mathbb{1}^{\otimes (\lambda_1-1)}\otimes  C \otimes\mathbb{1}^{\otimes (N-\lambda_n)}$, where $n=|\lambda|$ vary with the considered geometry and $C\in \mathcal{C}_{n,d}$. To simplify notation, and assuming this dependence on geometry is understood, we will use the shorthand $C_\lambda\in \mathcal{C}_{n,d}$.}. 
In essence, the collection $\mathcal{A}=\{\Lambda_s\}_{s=0}^t$ specifies the dynamical architecture of the circuit.
For example, the two-qudit brickwork circuit architecture we will consider later alternates  
$\Lambda_{2s} = \{(1,2), (3,4), \dots, (N-1,N)\}$ at even times with  $\Lambda_{2s+1} = \{(2,3), (4,5), \dots, (N-2,N-1)\}$ at odd ones, with $N$ an even number. 

For a given architecture $\mathcal{A}=\{\Lambda_s\}_{s=0}^t$, random clifford circuits are defined by independently sampling each local gate $C_\lambda$ uniformly from the finite set of Clifford gates acting on the set of sites $\lambda$. We are interested in computing the anticoncentration properties of random Clifford circuits. 
A central interest in our work are the \textit{$k$-replica expectation values}~\cite{Braccia2024computing,turkeshi2024hilbert}
\begin{equation}
    \mathfrak{O}_t\equiv \mathbb{E}_{\mathbf{C}_t}\llangle O | (\mathbf{C}_t\otimes \mathbf{C}_t^*)^{\otimes k} | \rho_0^{\otimes k}\rrangle\;,
\end{equation}
for a certain replica boundary states $|O\rrangle$ that specifies the observable of interest, while $|\rho_0^{\otimes k}\rrangle$ is the initial state of interest. 
(In the following $O$ will be a statistical proxy of anticoncentration.) 
The technology behind the computation of $\mathfrak{O}_t$ constitutes the so-called Clifford Weingarten calculus~\cite{leonetoappear,turkeshi2025magicspreadingrandomquantum,gross2021schurweyl}. 
In fact, by linearity of the expectation value and the statistical independence of the Clifford gates $C_\lambda$, the central building block is the replica transfer matrix, or moment operator,
\begin{equation}
    \mathcal{T}_\lambda = \mathbb{E}_{C_\lambda\in \mathcal{C}_{n,d}}[({C}_\lambda \otimes {C}^*_\lambda)^{\otimes k}]\;,\label{eq:transfer}
\end{equation}
with $n=|\lambda|$ is the number of qudits where the action of $C_\lambda$ is non-trivial. 

The replica transfer matrix is given in terms of the $k$-th Clifford commutant, namely the set of operators $A$ acting on $\mathcal{H}_{n,d}^{\otimes k}$ such that $[A,C^{\otimes k}_\lambda] = 0$ for all $C_\lambda \in \mathcal{C}_{n,d}$ that we denote by $\mathrm{Comm}_k(\mathcal{C}_{n,d})$.
Since $\mathcal{C}_{n,d} \subset \mathcal{U}(d^n)$, one has that $\mathrm{Comm}_k(\mathcal{C}_{n,d}) \supset \mathrm{Comm}_k(\mathcal{U}(d^n))$, 
where $\mathrm{Comm}_k(\mathcal{U}(d^n))=\mathrm{span}(\{A_\pi\}_{\pi\in \mathcal{S}_k})$ is spanned by the representation on $\mathcal{H}_{n,d}^{\otimes k}$ of the permutation group with $k$ elements $\mathcal{S}_k$~\cite{collins2006integration, mele2024introduction}. 
To identify the entire content of $\mathrm{Comm}_k(\mathcal{C}_{n,d})$, one uses that
the Clifford commutants are indexed by the stochastic Lagrangian subspaces of $\mathbb{Z}_d^{2k}$~\cite{gross2021schurweyl,montealegremora2020rankdeficientrepresentationsthetacorrespondence,nezami2020multipartite,roth2018recovering,haferkamp2022}. 
These are subspaces $\sigma\subset \mathbb{Z}_d^{2k}$ that satisfy the following conditions: (i) for every $(\pmb{x},\pmb{y})\in \sigma$ it holds $\sum_{i=0}^{k-1} (x_i^2 - y_i^2) = 0 \,\mathrm{mod}\,D$, where $D=4$ for qubits and $D=d$ for $d\ge 3$ prime, (ii) $\sigma$ has dimension $k$ as a vector space, (iii) the vector $\pmb{1}_{2k}=(1,1,\dots,1)$ is an element of $\sigma$. 
We denote the set of all such stochastic Lagrangian subspaces as $\Sigma_k(d)$, which has cardinality
\begin{equation}
|\Sigma_k(d)| = \prod_{m=0}^{k-2}(d^m+1)     = (-1;d)_{k-1}
\label{eq:sizesigma}
\end{equation}
where ${(a;\xi)_n \equiv \prod_{m=0}^{n-1}(1-a\xi^m)}$ is the q-Pochhammer symbol~\cite{Kac2002}.  
For any given stochastic Lagrangian subspace $\sigma$ we associate the corresponding vector $|\sigma\rrangle =\sum_{(\pmb{x},\pmb{y})\in \sigma}|\pmb{x},\pmb{y}\rrangle$ on an individual replica qudit $\mathcal{H}_d^{\otimes 2k}$. 
This elementary ingredient allows for constructing the action on $k$ copies of $n$ qudits via  $|\sigma\rrangle_\lambda \equiv \bigotimes_{i\in \lambda} |\sigma_i\rrangle$, where $\lambda$ are the qudits where $C_\lambda$ act non-trivially. 
The vectors $|\sigma\rrangle$ are linearly independent when $n = |\lambda| \geq k-1$; in that case, $|\mathrm{Comm}_k(\mathcal{C}_{n,d})| = |\Sigma_k(d)|$. Otherwise, some of these vectors are linearly dependent. 

The Schur-Weyl duality for Clifford unitaries expresses Eq.~\eqref{eq:transfer} in terms of the commutant $\mathrm{Comm}_k(\mathcal{C}_{n,d})$
\begin{equation}
    \mathcal{T}_\lambda = \sum_{\pi,\sigma \in \Sigma_{k}(d) } \mathrm{Wg}_{\pi,\sigma}(d^n) |\pi\rrangle_\lambda\llangle \sigma|_\lambda\;,
\end{equation}
where $\mathrm{Wg}(d^n)$ is the pseudo-inverse of the Gram matrix $\mathrm{G}_{\sigma,\pi} = \prod_{i=1}^n \llangle \sigma_i|\pi_i\rrangle$. 
To lighten the notation, we will employ a graphical notation in the following section. For instance, $\mathcal{T}_\lambda$ acting on two neighboring qudits is given by 
\begin{equation}
   \vcenter{\hbox{\includegraphics[width=0.65\linewidth]{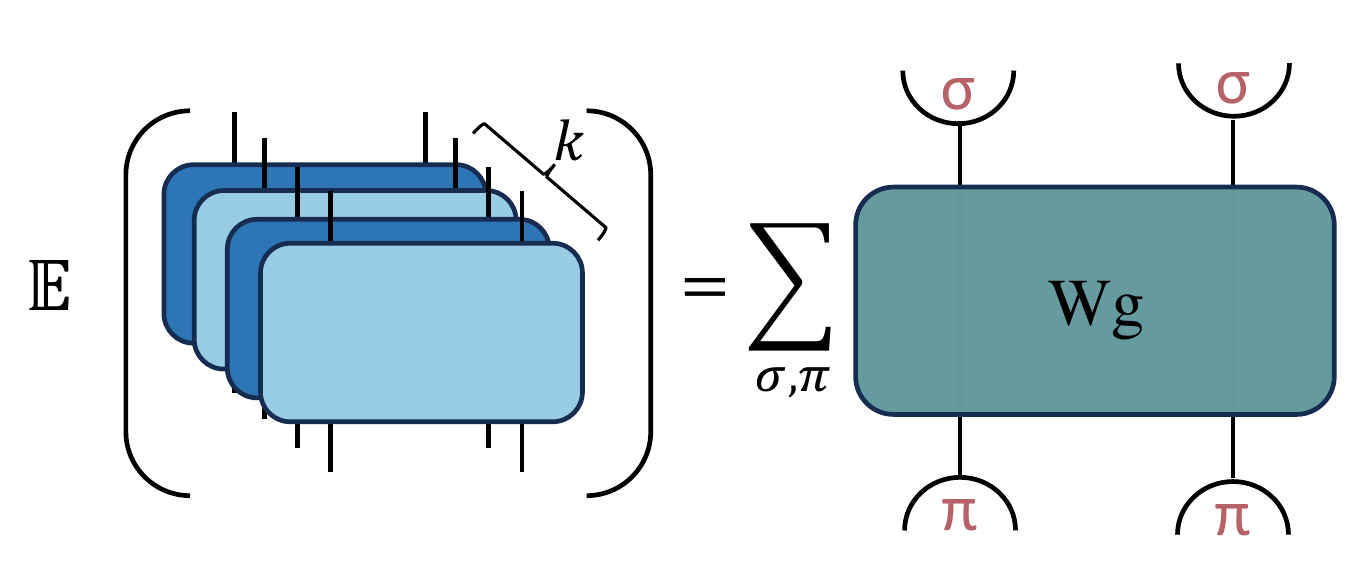}}}
    \label{fig:transfergate}
\end{equation}
whereas we will depict the gram matrix $\mathrm{G}$ as
\begin{equation}
    \vcenter{\hbox{\includegraphics[width=0.2\linewidth]{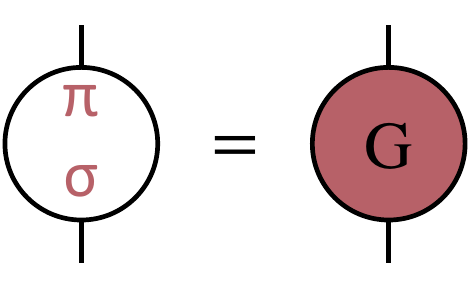}}}\;.
    \label{fig:Gram}
\end{equation} 
A fundamental property of the Gram and Weingarten matrix is that~\cite{gross2021schurweyl}
\begin{equation} 
    \sum_{\pi \in \Sigma_k(d)}\mathrm{G}_{\pi,\sigma}(d^n) =
    d^{k n} (-d^{-n};d)_{k-1} \coloneqq \mathcal{G}_{n,k, d}
    \label{eq:gramsum}
\end{equation}
and similarly 
\begin{equation}
    \sum_{\pi \in \Sigma_k(d)}\mathrm{Wg}_{\pi,\sigma}(d^n)=\frac{d^{-kn}}{(-d^{-n};d)_{k-1}} \coloneqq \mathcal{G}_{n,k, d}^{-1}\;.
    \label{eq:Wsum}
\end{equation}
It is important to stress that by increasing the number of qudits $n$ we can obtain arbitrarily large Clifford operators. However, the size \eqref{eq:sizesigma} of the commutant $|\Sigma_k(d)|$ for $k$ replicas explicitly depends on the dimension $d$ of each qudit. This contrasts with the usual Haar case, where, for $n\ge k-1$, commutant has size $k!$.

\section{Results}
\label{sec:results}
\subsection{Distribution of overlaps for random stabilizer states}
\label{subsec:resultsa}
We start our discussion by deriving the overlap distribution for random stabilizer states, obtained by averaging the action of Clifford Haar transformation onto $|\pmb{0}\rangle \to C \ket{\pmb{0}} \equiv \ket\Psi$. 
A stabilizer state is fully characterized in the computational basis $\mathcal{B}\equiv \{ |\pmb{x}\rangle\;|\; \pmb{x}\in \mathbb{Z}_d^N\}$ by a set of $d^g$ states $\mathcal{B}_\psi\equiv \{ \pmb{x}_j\}_{j=0,\dots,d^g-1}$ and phases  $\{ \varphi_j\}_{j=0,\dots,d^g-1}$ as 
\begin{equation}
    |\Psi\rangle = \frac{1}{\sqrt{d^g}} \sum_{i=0}^{d^g-1} \omega^{\varphi_i} |\pmb{x}_i\rangle\;, 
    \label{eq:psid}
\end{equation}
for a certain $g=0,\dots,N$. 
Therefore a fixed stabilizer state $|\Psi\rangle$ is uniformly spread over $d^g$ states, so that the rescaled overlap $w_{\pmb{y}}=d^N|\langle \pmb{y}|\Psi\rangle|^2= 1_{\mathcal{B}_\psi}(\pmb{y}) d^{N-g}$~\cite{sierant2022universal,turkeshi2023measuring} where $1_{\mathcal{A}}$ is the characteristic function of $\mathcal{A}$. 
As a result, the participation entropy is independent of the R\'enyi index 
$S[\ket{\Psi}] = S_k\equiv (1-k)^{-1}\log_d[I_k] = g$, where $I_k[|\Psi\rangle] \equiv\sum_{\pmb{x}=0}^{d^N-1} |\langle \pmb{x}|\Psi\rangle|^{2k}$ are the inverse participation ratios~\cite{luitz2014participation,liu2024quantum}.
Thus, we can identify $g$ with the participation entropy, which can be expressed in terms of the stabilizer tableau as~\cite{sierant2022universal}
\begin{equation}
    g= S= \mathrm{rk}_{\mathbb{Z}_d}(\mathsf{X})\label{eq:galois}
\end{equation}
with $\mathrm{rk}_{\mathbb{Z}_d}$ denotes the rank in the Galois field $\mathbb{Z}_d$. 
We are interested in the distribution of overlaps over random stabilizer states, namely
\begin{equation}
\begin{split}
    P(w) &= \mathbb{E}_{C}[\mathbb{E}_{\mathbf{y}}[\delta(w - w_{\pmb{y}})]]\\
    &=\mathbb{E}_{C}[\delta(w-d^N|\langle \pmb{0}|C|\pmb{0}\rangle|^2)]\;.
\end{split}
\end{equation}
Here, in the second step, we used Clifford invariance to reabsorb $\pmb{y}\mapsto \pmb{0}$. 

It is important to underscore the subtle difference between Hilbert space delocalization and anticoncentration: the former describes a property of an individual state, while the latter characterizes an ensemble of states. 
Specifically, the distribution of the overlap on 
a given realization $\mathbb{E}_{\pmb{y}}[\delta(w - w_{\pmb{y}})] = (1 - d^{g-N})\delta(w) + d^{g-N}\delta(w - d^{N-g})$
of a random stabilizer state is just made of two $\delta$ peaks and is not representative of the whole ensemble. 
Nonetheless, once averaging over the whole Clifford unitaries $\mathbb{E}_C[\ldots]$ a non-trivial form emerges and the previous discussion clarifies that it is determined by the statistics of $g$ and is agnostic to the phases. 
This is in sharp contrast with generic Haar unitaries, where at large $N$, a single realization is enough to extract the distribution of $w$ by varying $\pmb{y}$.

We extract the overlap distribution proceeding in two complementary ways. 
First, one can directly compute the expectation value $\mathbb{E}[w^{k}]$ via the replica transfer matrix.  As $\llangle 0,0|^{\otimes k}\cdot |\pi\rrangle=1$  for all $\pi\in \Sigma_k(d)$, we simply have
\begin{equation}
    \mathbb{E}[|\braket{\mathbf{0}|C|\mathbf{0}}|^{2k}] = \sum_{\sigma, \pi \in \Sigma(d)} \mathrm{Wg}_{\pi, \sigma}(d^N) = \frac{ |\Sigma_k(d)|}{\mathcal{G}_{N, k, d}}
    \;,
\label{eq:momWG}
\end{equation} 
where we used Eq.~\eqref{eq:Wsum} and the size of the commutant. However, reconstructing the distribution $P(w)$ from its moment is not obvious. 
To obtain it more directly, we observe that each stabilizer state is equivalent to its tableau representation \cite{ gottesman1997stabilizer, gottesman1998theory, nielsen2010quantum} and crucially, $g$ corresponds to the rank of $\mathsf{X}$~\cite{sierant2022universal}. Thus, the probability that a random stabilizer state has participation entropy $g$ is a counting problem of how many stabilizer states have the tableau $\mathsf{X}$ of rank $g$ in the Galois field $\mathbb{Z}_d$. 
{We note that row operations on the matrix $\mathsf{S}$ lead to the same state. As a result, the number of degeneracies for the tableau $\mathsf{X}$ is equivalent to the number of echelon matrices with $g$ rows over $N$ terms, given by the q-Binomial $\binom{N}{g}_d$~\cite{Kac2002}. 
Once fixed the structure of $\mathsf{X}$, there is an additional degeneracy on setting the terms in $\mathsf{Z}$ which, modulo Gauss row operations, is the possible non-trivial vectors in $\mathsf{Z}$ over the non-trivial rows in $\mathsf{X}$, given by $d^{g(g+1)/2}$. Finally, an additional degeneracy $d^N$ comes from the phases $\phi$. 
In summary, the number of stabilizer states with participation entropy $g$ is given by 
\begin{equation}
    \aleph_g \equiv \# \{
    C \in \mathcal{C}_{N,d} |
S[C\ket{\pmb{0}}] = g\} = d^N d^{g(g+1)/2} \binom{N}{g}_d\;.
\end{equation}
As a simple consistency check, we note that $\sum_{g=0}^N \aleph_g = d^N(-d;d)_N = |\mathrm{Stab}_{N,d}|$ reproduces the number of all stabilizer states on $N$ qudits. 
Thus, the distribution of the participation entropy $g$ is simply given by
\begin{equation}
    \mathcal{P}_\mathrm{Haar}(g) = \frac{\aleph_g}{|\mathrm{Stab}_{N,d}|}=\binom{N}{g}_d\frac{d^{g(g+1)/2}}{(-d;d)_N}\;. 
    \label{eq:HaarCdistribution}
\end{equation}
This expression is one of the main results of our work: it is the analogue to the Porter-Thomas distribution of the Haar unitary ensemble, which we dub \textit{Clifford-Porter-Thomas distribution}.

An important consequence of Eq.~\eqref{eq:psid} is the loss of self-averaging in the Clifford ensemble. 
To illustrate this point, consider the quenched and annealed averages of the participation entropy over an ensemble $\mathcal{E}$, respectively defined by $\overline{S}_k\equiv\mathbb{E}_\mathcal{E}[S_k]$ and $\tilde{S}_k\equiv (1-k)^{-1}\log_d \mathbb{E}_\mathcal{E}[I_k]$. 
Self-averaging occurs when these two quantities are equal, up to exponentially small corrections in system size $N$. 
This holds for Haar-random unitary transformations $\mathcal{U}(d^N)$, where measure concentration ensures that for large systems ($N\gg1$), a single realization of a Haar-random unitary is typical. 
In other words, the probability that any polynomial function $p(U,U^\dagger)$ deviates significantly from its expectation value $\mathbb{E}[p(U,U^\dagger)]$ is exponentially small in $N$.
In contrast, for Clifford unitaries, we can compute the average inverse participation ratios 
using Eq.~\eqref{eq:HaarCdistribution} and the $q$-Binomial theorem
\begin{equation}
\begin{split}
    I_k^\mathrm{Haar}&\equiv\sum_{g=0}^{N} d^{(1-k)g} 
\mathcal{P}_{\rm Haar}(g)
= 
\frac{\left(-d^{2-k};d\right){}_N}{(-d;d)_N}
\;\\
&= 
\frac{d^{(1-k)N} (-1;d)_{k-1}}{\left(-d^{-N};d\right){}_{k-1}}
\;,
    \label{eq:IPRHaarCl}
\end{split}
\end{equation} 
where the second line confirms the agreement with Eq.~\eqref{eq:momWG}.
This inverse participation ratios of the Clifford ensemble in Eq.~\eqref{eq:IPRHaarCl} exhibit multifractal behavior~\cite{mace2019multifractal, backer2019multifractal}, meaning they depend non-trivially on the R\'enyi index $k$. In particular, the annealed average participation entropy over the Clifford group is given by 
\begin{equation}
    \tilde{S}_k =  \frac{1}{1-k}\log_d[\mathbb{E}[I_k^\mathrm{Haar}]] = N + c_k + O(e^{-\gamma N}) \;,
    \label{eq:annealed}
\end{equation}
where $c_k = (1-k)^{-1} \sum_{m=0}^{k-2} \log_d(1+d^m)$ scales as $c_k\sim -k/2$ at large $k\gg1$. 
This contrasts the average participation entropy~\cite{sierant2022universal}
\begin{equation}
    \overline{S} = \sum_{g=0}^N g \mathcal{P}_\mathrm{Haar}(g) =  N + c + O(e^{-\gamma N})\;,
    \label{eq:Squench}
\end{equation}
where $c=-\sum_{g=1}^{\infty} (d^g+1)^{-1}$, which is independent of the R\'enyi index and thus follows a fractal behavior~\cite{turkeshi2023measuring}. The discrepancy between annealed \eqref{eq:annealed} and quenched average \eqref{eq:Squench} quantifies how Clifford circuits are not self-averaging.

Finally, we emphasize that the explicit distribution $\mathcal{P}_\mathrm{Haar}(g)$ enables us to compute the \emph{exact} higher order moments of the participation entropy for stabilizer states. This is a significant advantage over Haar-random cases, where such calculations require resumming an intricate logarithmic expansion, cf. for instance~\cite{zhou2020entanglement,zhou2019emergent}.

The leading factor in Eq.\eqref{eq:annealed} and \eqref{eq:Squench} indicates that the distribution of $g$ concentrates around its maximal value. We can thus explore the thermodynamic limit $N\to \infty$ defining $n=N-g\in \{0,1,\dots,N\}$, with the limiting distribution 
\begin{equation}
\label{eq:Haarcliff}
\mathcal{P}_\mathrm{Haar}^{\infty}(n) = \frac{1}{(-d^{-1}; d^{-1})_\infty}\frac{d^{-n(n+1)/2}}{(d^{-1}; d^{-1})_n} \;.
\end{equation}
This expression is the scaling limit of the Clifford-Porter-Thomas, similarly to the exponential distribution being the scaling limit of the Haar unitary Porter-Thomas~\footnote{Performing a change of variable, we can express the overlap distribution in terms of the overlap $\mathcal{w}= d^{N-g}$, the result is an exponential with a polylog in $\mathcal{w}$ at the exponent.}, cf. App.~\ref{app:A}. 
We can use this result to recover the moments of the overlaps using $\mathbb{E}[w^k | n] = d^{n (k-1)} $. 
Summing over the $n$, we obtain
\begin{equation}
\mathbb{E}[w^k] =
\sum_n \mathbb{E}[w^k | n]
\mathcal{P}_\mathrm{Haar}^{\infty}(n) = (-d^{k-2}; d^{-1})_{k-1} 
\end{equation}
which, up to an overall prefactor $d^{(1-k)N}$, exactly reproduces Eq.~\eqref{eq:IPRHaarCl} in the large $N$ limit.  
As a numerical verification, we compute Eq.~\eqref{eq:galois} for $\mathcal{N}=2\times 10^7$  random stabilizer states across different systems sizes and compare the resulting empirical distribution with the analytical prediction Eq.~\eqref{eq:Haarcliff}.  
For this purpose, we use the tableau simulator \textsc{STIM}~\cite{Gidney2021stimfaststabilizer} and the rank-revealing algorithm in Ref.~\cite{sierant2023entanglement}. Our results, presented in Fig.~\ref{fig:rmps} (a), demonstrate already small system sizes saturate the Haar prediction. 
In the next section, we investigate how structured dynamical systems approach the Haar scaling behavior of Eq.~\eqref{eq:IPRHaarCl}, focusing on random tensor network states and quantum circuits.

\subsection{Clifford random tensor network states}\label{subsec:resultsb}
The previous discussion presented the distribution of overlaps for random stabilizer states. 
We now inquire about the approach to Eqs.~\eqref{eq:HaarCdistribution} and \eqref{eq:IPRHaarCl} for the class of Clifford random matrix product states (CRMPS).  
We recall that matrix product states (MPS) are vectors in the Hilbert space whose wave-function is fixed by a sequence of $N$ tensors, denoted $\Gamma^{x_i}_{\alpha_i, \beta_i}$ where $\alpha,\;\beta\in\{0,1,\dots,\chi-1\}$ are the virtual indices living in a $\chi$-dimensional auxiliary space, and $x_i\in \{0,\dots,d-1\}$ is the qudit variable in computational basis~\cite{garnerone2010typicality, garnerone2010statistical, haferkamp2021emergent, haag2023typical, lami2024quantum, lami2025anticoncentration}. 
Clifford random matrix product states are obtained when the tensors $\Gamma$ are induced by Clifford transformations~\cite{li2024statistical}. Throughout this work, we fix $\chi=d^r$ for $r\in \{0,1,\dots,N-1\}$, so that the explicit wave-function is given by 
\begin{equation}
\begin{split}
    \langle \pmb{x}|\Psi\rangle & = \sum_{\alpha_1,\dots,\alpha_{N-r}=0}^{d^r-1} \sum_{\alpha_{N-r+1}=0}^{d^{r-1}-1}\cdots \sum_{\alpha_{N-1}=0}^{d-1} \\ &\qquad\qquad C_{0,0,x_1,\alpha_1}^{(1)}C_{0,\alpha_1,x_2,\alpha_2}^{(2)}\cdots C_{\alpha_{N-1},x_N}^{(N)}\;,
    \label{eq:psiMPS}
\end{split}
\end{equation}
where each Clifford gate is uniformly drawn $C^{(j)}\in \mathcal{C}_{r+1,d}$ for $j\le N_r$ and is reshaped as a $d\times d^r\times d\times d^r$ tensor, while $C^{(N-r+i)}\in \mathcal{C}_{r-i+1,d}$ is reshaped into a $d^{r-i+1}\times d\times d^{r-i}$ tensor for $i=1,\dots,r$. We note that the decreasing bond dimensions ensure that the state is normalized in the right canonical form~\cite{schollwock2011the,itensor}. 
To lighten the notation, from now on, we will use the standard graphical representation of tensor networks~\cite{orus2014a}, inferring contractions from lines ending in a given geometrical object. 
For instance, a Clifford random matrix product state is represented by 
\begin{equation}\label{eq:rmps_bulk}
\begin{tikzpicture}[baseline=(current  bounding  box.center),scale=0.8]
\definecolor{mycolor}{rgb}{0.6, 0.8, 0.9}
    \foreach \x in {1,...,7}{
        \draw[thick, black] (1.1*\x,0) -- (1.1*\x,1.6);
        \draw[line width=0.4mm, fill=white] (1.1*\x,0) circle (0.2);
    }
    \draw[line width=0.8mm, blue!60!black, dotted] (-0.3,0.8) -- (0.2,0.8);
    \draw[line width=0.8mm, blue!60!black] (0.2,0.8) -- (8.4,0.8);
    \draw[line width=0.8mm, blue!60!black, dotted] (8.4,0.8) -- (8.9,0.8);
    \foreach \x in {1,...,7}{
        \draw[line width=0.5mm, fill=mycolor, rounded corners=5pt] (1.1*\x-0.4,0.4) rectangle (1.1*\x+0.4,1.2);
        \pgfmathsetmacro{\y}{int(\x - 3)}
    }
\end{tikzpicture}\;,
\end{equation}
with the open thin indices the uncontracted physical dimensions, think blue indices denote the virtual indices, while the white circles represent the $|0\rangle$ states. 

We are interested in the overlap distribution over the ensemble of Clifford random matrix product states. 
For this scope, we compute the inverse participation ratios
\begin{equation}
    I_k^\mathrm{CRMPS} = \mathbb{E}_{\rm CRMPS}\left[\sum_{\pmb{x}=0}^{d^N-1} |\langle\pmb{x} | \Psi\rangle|^{2k}\right]= d^{(1-k)N}\mathbb{E}\left[w^k\right]\;,
    \label{eq:IPR}
\end{equation}
where we used the local Clifford invariance to rephrase the problem in terms of $w=d^N |\langle \pmb{0}|\Psi\rangle|^2$, where $|\Psi\rangle$ is an MPS as in Eq.~\eqref{eq:psiMPS} and average is taken over the realizations of the gates. 
By reshaping the tensor network, this overlap is computed from the staircase quantum circuit consisting of $N-r$ sequential Cliffords $C^{(i)}$ on $r+1$ qudits all initialized in $|0\rangle$\footnote{
This circuit is specified by Eq.~\eqref{eq:circuit} with $\Lambda_s =\{ \{s+1,\dots,s+r+1\}\}$ for $s=0,\dots,N-r-1$}. The resulting amplitude is 
\begin{equation}\label{eq:rmpsOBC}
   \langle\pmb{0} | \mathbf{C}|\pmb{0}\rangle=\begin{tikzpicture}[baseline=(current bounding box.center),scale=0.6,rotate=270]
   \definecolor{mycolor}{rgb}{0.6,0.8,0.9} 
    \foreach \x in {1,...,6}{
        \draw[thick, black] (\x,0) -- (\x,8.5);
        \draw[line width=0.4mm, fill=white] (\x,0) circle (0.2);
        \draw[line width=0.4mm, fill=white] (\x,8.7) circle (0.2);
    }
    \foreach \x in {1,...,5}{
        \draw[line width=0.8mm, blue!60!black] (\x,1.5*\x-1.3) -- (\x,1.5*\x);
        \draw[line width=0.4mm, fill=mycolor, rounded corners=5pt] 
            (\x-0.2,1.5*\x-0.6) rectangle (\x+1.2,1.5*\x+0.2);
    }
    \draw[line width=0.8mm, blue!60!black] (6,1.5*6-1.3) -- (6,8.5);
\end{tikzpicture}\;,
\end{equation}
where again, blue contractions correspond to the virtual indices in $\chi=d^r$, which can be seen as resulting from the tensor product of $r$ qudits. 
\begin{figure}[t]
    \centering
    \includegraphics[width=\columnwidth]{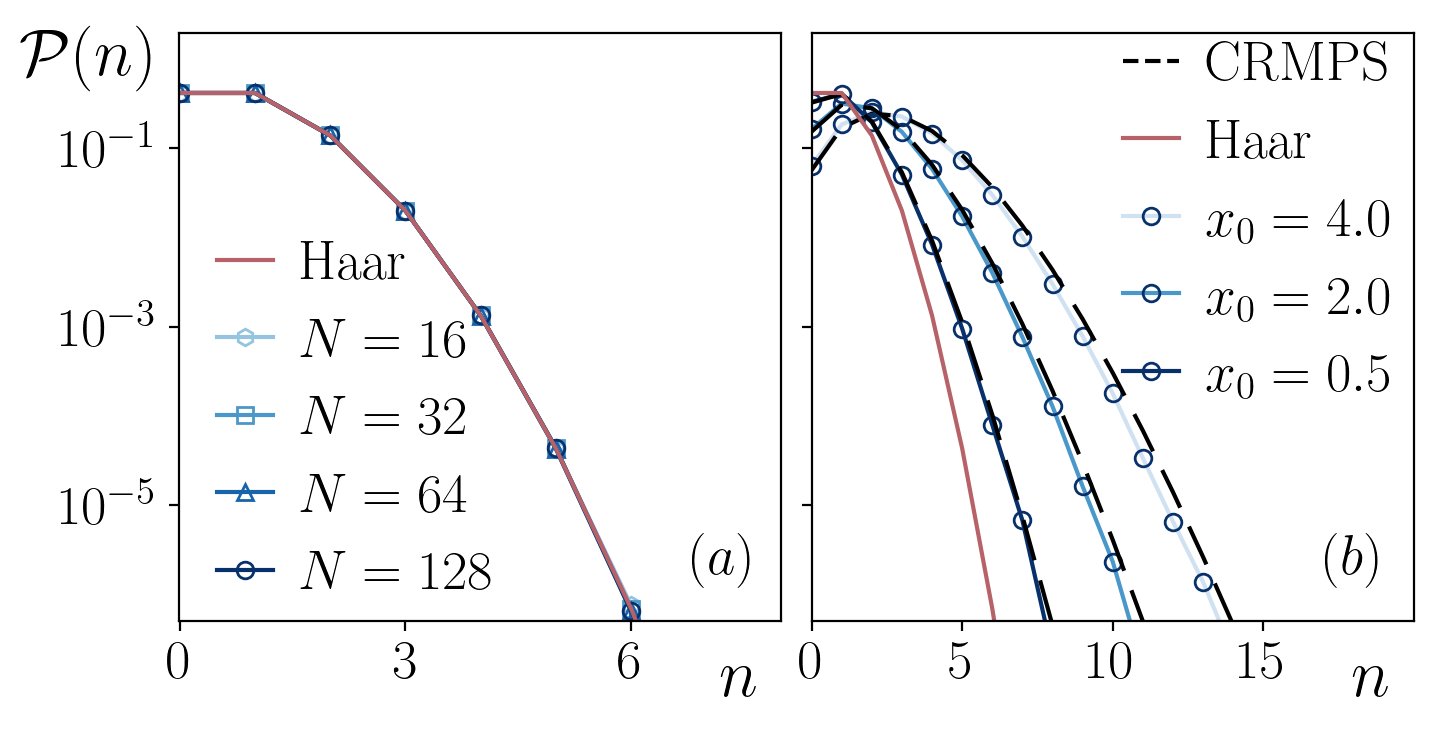}
    \caption{(a) Numerical sampling of random stabilizer states over $N$ qubits ($d=2$), compared to the Haar prediction (red line in the plot) in Eq.~\eqref{eq:IPRHaarCl}. 
    The data is in quantitative agreement with our analytical predictions. 
    (b) Numerical sampling of the overlap distribution of Clifford random matrix product states for $N=256$ varying $x=N/\chi$. The dashed black line is the analytical prediction Eq.~\eqref{eq:IPRrandomMPS} which is in robust agreement with our data. For reference, we plot the Clifford Haar distribution Eq.~\eqref{eq:IPRHaarCl} in red, highlighting the substantial difference in the scaling regime $N,\chi\to\infty$ with $x_0=N/\chi$ fixed. 
    In both cases, we consider $\mathcal{N}=2\times 10^7$ circuit realizations. 
    }
    \label{fig:rmps}
\end{figure}
Performing the replica trick and the Clifford averages, we obtain 
\begin{equation}\label{eq:rmpsOBC1}
   \mathbb{E}\left[w^k\right]=d^{k N}\;\begin{tikzpicture}[baseline=(current bounding box.center),scale=0.55,rotate=270]
   \definecolor{mycolor}{rgb}{0.392,0.6,0.616} 
   \definecolor{mycolor2}{rgb}{0.717,0.38,0.407} 
    \foreach \x in {1,...,6}{
        \draw[thick, black] (\x,0) -- (\x,8.5);
        \draw[line width=0.4mm, fill=black] (\x,0) circle (0.2);
        \draw[line width=0.4mm, fill=black] (\x,8.7) circle (0.2);
    }
    
    \foreach \x in {1,...,5}{
        \def\yA{1.5*\x-1.3}
        \def\yB{1.5*\x}
        \draw[line width=0.8mm, blue!60!black] (\x,\yA) -- (\x,\yB);
        \draw[line width=0.4mm, fill=mycolor, rounded corners=5pt] (\x-0.2,1.5*\x-0.6) rectangle (\x+1.2,1.5*\x+0.2);
    }
    \foreach \x in {2,...,5}{
        \def\yA{1.5*\x-1.3};
        \def\yB{1.5*\x};
        \def\zz{1.5*\x-0.95};
        \draw[line width=0.4mm, fill=mycolor2] (\x, \zz) circle (0.2);
    }
    \draw[line width=0.8mm, blue!60!black] (6,1.5*6-1.3) -- (6,8.5);
\end{tikzpicture}\;,
\end{equation}
where the contractions with the black dot are with the on-site replica $|0,0\rrangle^{\otimes k}$ state, and we inserted the transfer matrix and the Gram contractions in Eqs.~\eqref{fig:transfergate} and \eqref{fig:Gram}. 
Noticing that $\llangle 0,0|^{\otimes k}\cdot |\pi\rrangle=1$  for all $\pi\in \Sigma_k(d)$ we can iteratively resum this circuit using Eqs.~\eqref{eq:gramsum} and \eqref{eq:Wsum} of the Weingarten calculus, leading to the final expression 
\begin{equation}
\begin{split}
    I_k^{\mathrm{CRMPS}} 
    &= d^N \frac{\mathcal{G}_{r,k,d}^{N - r - 1}}{\mathcal{G}_{r+1, k, d}^{N-r}} |\Sigma_k(d)|\\
    &= I_k^{\mathrm{Haar}}(-d^{-N};d)_{k-1}\frac{\left[(-d^{-r};d)_{k-1}\right]^{N-r-1}}{\left[(-d^{-1-r};d)_{k-1}\right]^{N-r}}\;.
    \label{eq:IPRrandomMPS}
\end{split}
\end{equation}
As explained in \cite{christopoulos2024universal, lami2025anticoncentration},
universality is expected to emerge in the thermodynamic limit $N \to \infty$ for quantum circuits with large depth, see also Refs.~\cite{PhysRevLett.130.140403, Chan2022, deluca2024universalityclassespurificationnonunitary}. 
In the case of random MPS, this amounts to consider the scaling limit $N,\chi\to\infty$, with $x_0=N/\chi$ constant. Using the q-analog of the Stirling approximation and exponentiating we have, in the scaling limit
\begin{equation}
    I_k^{\mathrm{CRMPS}} = I_k^{\mathrm{Haar}} \exp\left[\frac{d^k-d}{d^2}x\right]\;,
    \label{eq:scalingCRMPS}
\end{equation}
where the scaling variable is defined up to the second order in $O(1/N)$ by 
\begin{equation}
    x\simeq x_0 \left[ 1- \frac{\log_d(N/x_0)}{N} - \frac{d}{(d-1)N} \right]+ O(x_0^2) \;.\label{eq:ziioooo}
\end{equation} 
Interestingly, when $d=2$, Eq.~\eqref{eq:IPRrandomMPS} recovers, up to $k=2,3$, the results for Haar unitary random MPS \cite{lami2025anticoncentration}, as well as for generic $d\ge 3$ prime and $k=2$, consistently with the state-design properties of the Clifford group \cite{zhu2016cliffordgroupfailsgracefully, kueng2015qubitstabilizerstatescomplex,webb2016cliffordgroupformsunitary}. 
We recall that each stabilizer state is spread over a set of $d^g$ states (Eq.~\eqref{eq:psid}). In the limit $N \to \infty$, the variable $g$ concentrates around its maximum and we are interested in the distribution of $n = N - g$ over the realizations of the CRMPS in the scaling limit. Since $I_k = d^{N(1-k)} \mathbb{E}_n [d^{-n(1-k)} ]$, we can read the characteristic function of the random variable $n$ from Eq.~\eqref{eq:scalingCRMPS}. Let us denote as $n_0$ the variable associated with the Haar distribution in Eq.~\eqref{eq:Haarcliff}. Then, Eq.~\eqref{eq:scalingCRMPS} is consistent with $n=n_0 + p$ where $n_0$ and $p$ are drawn independently and $p$ follows the Poisson distribution, i.e.
\begin{equation}
    \mathcal{P}(p) = \frac{\lambda^p e^{-\lambda}}{p!} \;, \quad \lambda = \frac{x}{d}
\end{equation}
By using simple convolution, we can recover the distribution of $n$ for Clifford random matrix product states
\begin{equation}
    \mathcal{P}(n) = \frac{e^{- x/d} d^{-n} x^n}{(-d^{-1};d^{-1})_{\infty }} \sum _{p=0}^n \frac{x^{-p} d^{-p(p-1)/2}}{(n-p)! (d^{-1};d^{-1})_{p}}
    \;.\label{eq:ziopera}
\end{equation}
The above expression is a central result of our work. 
It reveals that the statistics of random Clifford tensor network combines the features of random stabilizer states~\eqref{eq:HaarCdistribution}, with Poissonian corrections in the scaling variable $x$. As we shall see, in agreement with the aforementioned universality, this result holds for much more general classes of systems, with $x$ a (single) fitting parameter. 

We conclude this section by benchmarking our analytical prediction with numerical sampling for qubit systems ($d=2$). 
Using standard tableau simulator~\cite{Gidney2021stimfaststabilizer}, we compute Eq.~\eqref{eq:rmpsOBC} for $N=256$ qubits and varying $x_0=N/\chi$. 
Our results are presented in Fig.~\ref{fig:rmps}(b). We see that our analytical prediction quantitatively describes the data for $\mathcal{N}=2\times 10^7$ disorder realizations~\footnote{
We have checked, but not present here, that analogous results hold for other values of $N$. The interested reader can access our complete datasets and code in Ref.~\cite{data}.
}.

\subsection{Anticoncentration in Clifford circuits}

\subsubsection{From random tensor networks to shallow circuits}
Even though our results for RTN provide a solid analytical foundation, they do not fully capture the local and dynamical evolution of more general quantum systems. 
To address this, we extend our analysis in this section to quantum circuits, employing both numerical and analytical methods. 
Specifically, we consider local random quantum circuits, where each gate—sampled from the Clifford ensemble—acts on neighboring sites in a brickwork geometry. 
In this setup, each discrete time step corresponds to an extensive number of gates acting on the system. Notably, a brickwall circuit with $O(N)$ layers -- hence $O(N^2)$ local gates, is sufficient to generate a global Clifford transformation~\cite{Bravyi_2021}. 
We now address the key question of whether anticoncentration occurs at linear depth in system size or if this cost can be further reduced.

First, we focus on a case that has recently gained attention in the context of Haar-random unitary gates: ``glued'' shallow circuits~\cite{schuster2025randomunitariesextremelylow}. 
This framework allows us to identify the timescale at which anticoncentration occurs in Clifford systems, similar to what happens for Haar unitary circuits.

Strictly speaking, the staircase circuit associated with the construction of CRMPS \eqref{eq:rmpsOBC} clearly has linear depth when $r=o(N)$. However, this geometry can be rearranged to obtain a shallow circuit with the same overlap distribution. To this end, we choose the physical dimension to match the bond dimension $\chi=d^r$. Rearranging the system, we recover a Clifford staircase circuit (CSC), but with each Clifford gate acting on two neighboring $\chi=d^r$ dimensional qudits. 
Repeating the analysis from the previous section, we now account for the fact that each random tensor acts only on the virtual bonds, with a total of $\chi^2 = d^{2r}$ sites. Additionally, given N as the total number of initial qudits, the number of gates in the system is $(N/r)-1$. Under these conditions, the inverse participation ratios are
\begin{equation}
\begin{split}
    I_k^{\mathrm{CSC}}  
    &= I_k^{\mathrm{Haar}}(-d^{-N};d)_{k-1}\frac{\left[(-d^{-r};d)_{k-1}\right]^{\frac{N-2r}{r}}}{\left[(-d^{-2r};d)_{k-1}\right]^{\frac{N-r}{r}}}.
\end{split}
\end{equation}
Nevertheless, the CSC can be reshaped to the Clifford glued circuit (GC) since, on average~\footnote{This follows from the fact that individual rows and columns of the Clifford gate $C$ exhibit the same statistical properties, akin to Haar-random gates.}
\begin{equation}
    \centering
    \vcenter{\hbox{\includegraphics[width=\columnwidth]{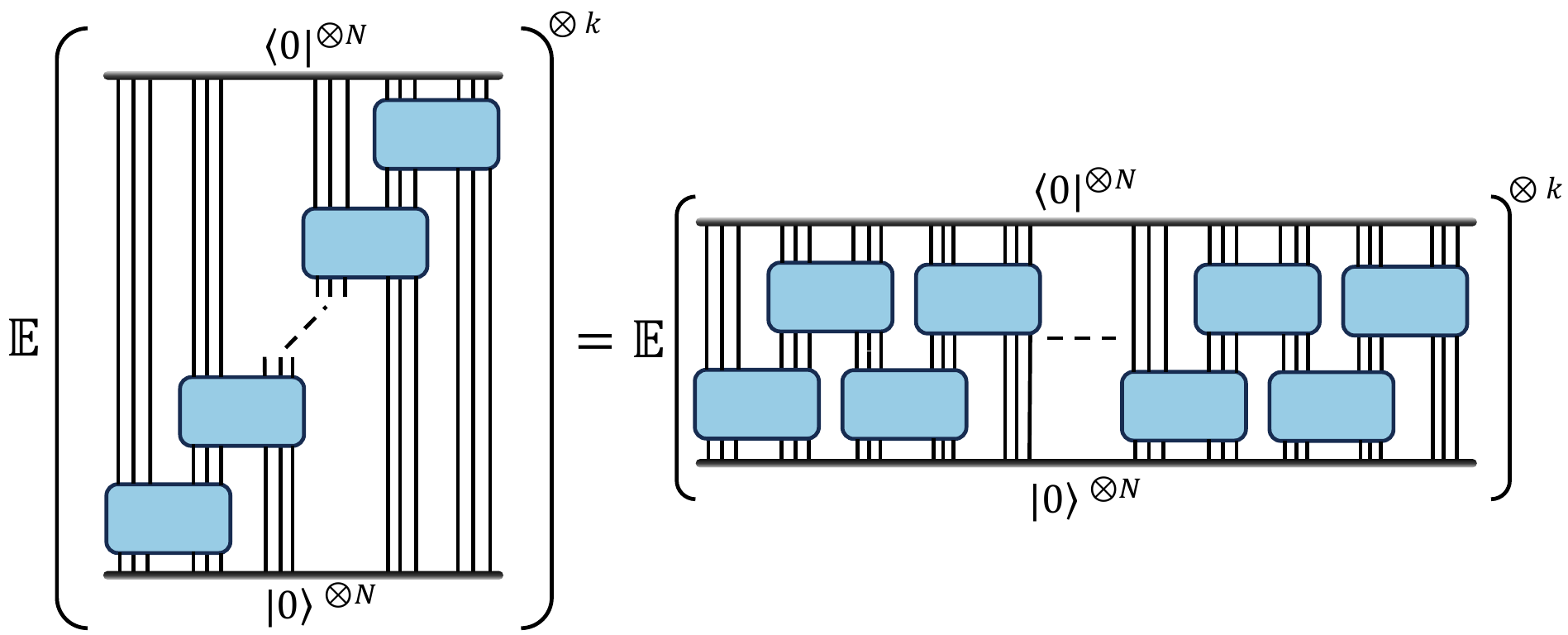}}}
    \label{fig:stitched_circuit}
\end{equation}
implying the same anticoncentration behavior. 
Taking the scaling limit $N, \chi \rightarrow \infty$ while keeping $x_0 = \frac{d}{d-1} \frac{N}{\log_d(\chi) \chi}$ constant, we obtain the scaling limit expression
\begin{equation}
    I_k^\mathrm{CSC}=I_k^\mathrm{GC} = I_k^\mathrm{Haar} \exp \left[\frac{d^k-d}{d^2} x \right]\;,
\end{equation}
recovering Eq.~\eqref{eq:scalingCRMPS} with $x \simeq  x_0 [1-2/(x_0 \chi)]$. 
Thus, setting $r \propto \log N$ is sufficient for anticoncentration toward random stabilizer states. Notably, since individual gates acting on $\chi^2 = d^{2r}$ can be implemented in linear depth in $2r$, this implies that local Clifford circuits anticoncentrate at depth $O(\log N)$.  
Moreover, Eq.~\eqref{eq:ziopera} remains valid, controlling fluctuations relative to the Haar-Clifford case.

We conclude this subsection with a heuristic argument: introducing interactions between gates across different patches would further accelerate anticoncentration. However, developing a rigorous mathematical framework to fully characterize this effect is significantly more challenging.
For this reason, in the following section, we will present numerical benchmarks, further corroborating our conclusions on Clifford circuits anticoncentration.

\subsubsection{Numerics}
To establish our numerical analysis, we begin by introducing the Clifford replica tensor network. We consider a brickwork circuit, as defined in Eq.~\eqref{eq:circuit}, where we distinguish between even and odd layers. Each two-qudit gate is an independent, identically distributed random Clifford gate acting on neighboring sites, given by:
\begin{equation}
\vcenter{\hbox{\includegraphics[width=0.4\columnwidth]{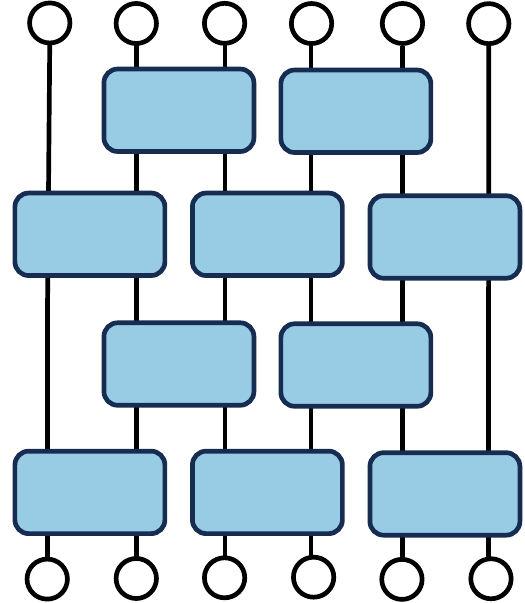}}}
    \label{fig:brickwall}
\end{equation}
Again, we are interested in Eq.~\eqref{eq:IPR} but for the ensemble of states generated by fixed depth Clifford brickwall circuits. 
Upon performing the average, we now have a tensor network whose degrees of freedom live in the commutant space of dimension $d_\mathrm{eff} = |\Sigma_k(d)|$. When the Clifford group forms a $k$ design, this replica tensor network exactly coincides with that of Haar circuits, i.e. $|\Sigma_k(d)| = k!$~\cite{turkeshi2024hilbert}. 
Beyond these limits, the Clifford commutant includes defect subspaces, enlarging substantially the local space dimension (see Eq.~\eqref{eq:sizesigma}).

Translating the circuit into replica space using the graphical convention defined in Sec.~\ref{sec:methods}, we obtain the following structure, where we distinguish between bulk gates and the initial boundary conditions
\begin{equation}
\vcenter{\hbox{\includegraphics[width=0.6\columnwidth]{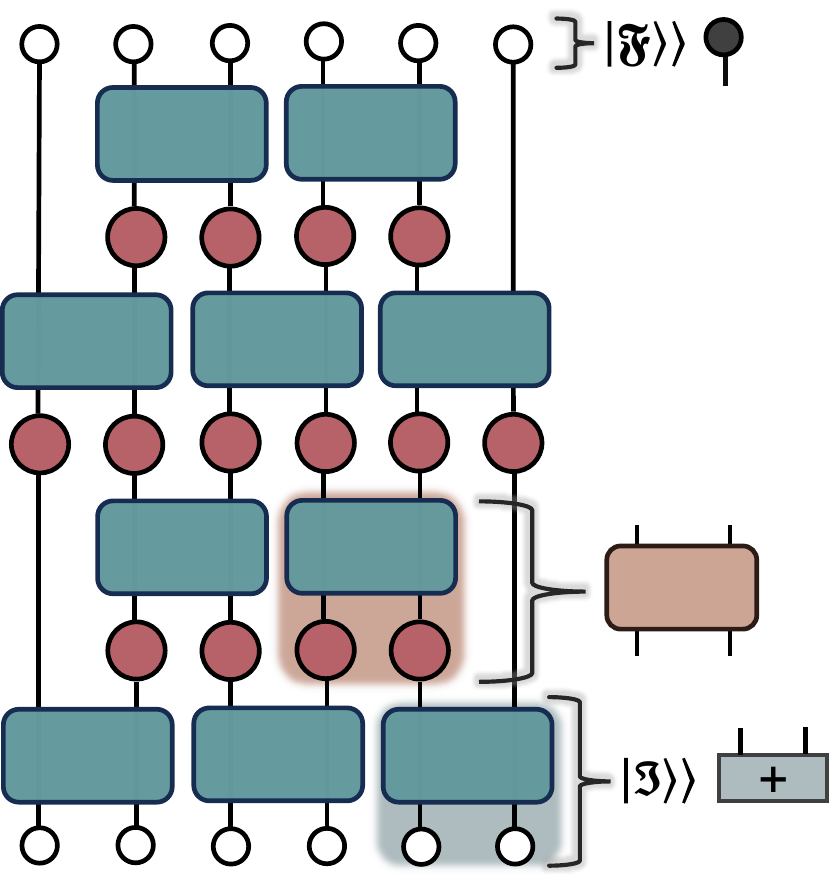}}}\label{fig:replica_brickwall}\;.
\end{equation}
The bulk gates (sienna color) are given by
\begin{equation}
\begin{split}
    \mathfrak{W}_{i,i+1} = &\sum_{\pi,\rho,\sigma,\tau}\mathrm{G}^{(i)}_{\sigma,\pi}(d) \mathrm{G}^{(i+1)}_{\rho,\pi}(d) \cdot \\ & \mathrm{Wg}^{(i,i+1)}_{\pi,\tau}(d^2)|\sigma_i\rho_{i+1}\rrangle\llangle \tilde{\pi}_i\tilde{\pi}_{i+1}|\;,
\end{split}
\end{equation}
where $\llangle \tilde{\pi}|\tau\rrangle = \delta_{\pi,\tau}$. The final contraction is simply $\llangle \mathfrak{F}|\equiv  \bigotimes_{i=1}^N(\sum_{\pi} \llangle \tilde{\pi}|)$ while the initial contraction is given by a tensor product $|\mathfrak{I}\rrangle\equiv \prod_{i=1}^{N/2} |+\rrangle_{2i-1,2i}$ with
\begin{equation}
\begin{split}
    |+\rrangle_{i,i+1} &\equiv \sum_{\sigma,\tau} \mathrm{Wg}_{\sigma,\tau}^{(i,i+1)}(d^2) |\sigma_i\sigma_{i+1}\rrangle\llangle \tau_{i}\tau_{i+1}|(|0,0\rrangle^{\otimes k}) \\ &= \frac{d^{2(1-k)}}{(d^{-2};d)_{k-1}}  \sum_{\sigma} |\sigma_i\sigma_{i+1}\rrangle\;,
    \end{split}
\end{equation}
replica Bell pairs. This leads to the contracted circuit 
\begin{equation}
\vcenter{\hbox{\includegraphics[width=0.5\linewidth]{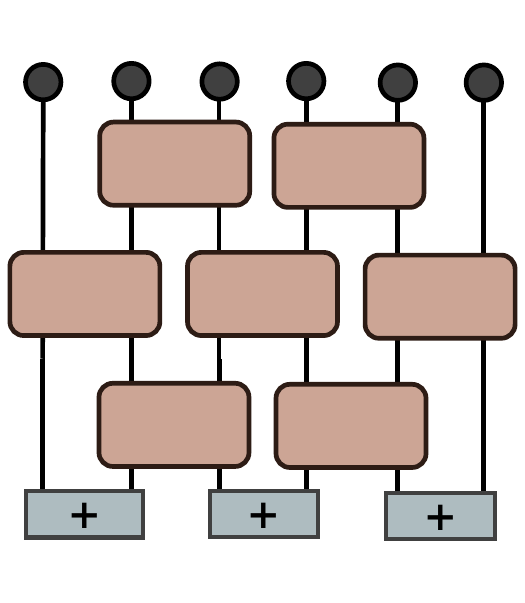}}}\label{fig:contracted_brickwall}
\end{equation}
which is at the base of our numerical implementation.

For simplicity, we consider a replica tensor network with  $d = 3 $ and $ d = 5 $, where  $k = 3$  copies of the system are sufficient to observe Clifford anticoncentration behavior beyond Porter-Thomas statistics.
With these specifications, we can efficiently perform large-scale numerical simulations up to $512$ qudits. This is possible because the bulk gates are non-unitary, preserving the replica bond dimension at $O(d_\mathrm{eff}^\mu)$ for some exponent $\mu$, where $d_\mathrm{eff}=2(d+1)$. 

To define the commutant required for constructing the bulk gates, we refer to \cite{gross2021schurweyl}. 
One could apply the general construction of the commutant given in Sec.~\ref{subsec:methodsb}, but for the cases of interest, we start from the fact that the commutant for $k=3$ always contains within it the group of permutations $\mathcal{S}_3$. Explicitly for $\sigma\in \mathcal{S}_3$, in the double-space representation, the operators $ \kket{\sigma_i}$, defined for each qudit $i$, are expressed in terms of the computational basis $\kket{\pmb{x}}$ as  
\begin{equation}
\bbra{x_1^i\bar{x}_1^{i}x_2^i\bar{x}_2^{i}x_3^i\bar{x}_3^{i}}\sigma_i\rrangle = \prod_{m=0}^2\delta_{x_m^i,\bar{x}_m^{\sigma(i)}}.
\end{equation}
To construct the Clifford commutant, we need to find extra elements to saturate the  dimension~\eqref{eq:sizesigma}.  Thus, we introduce the intrinsic Clifford commutant operators
\begin{equation}
    Q_d = \frac{1}{d} \sum_{P\in \mathcal{P}_{1,d}} P^{\otimes 2} \otimes P^{d-2},
\end{equation}
which induce scalable measures of magic resources or nonstabilizerness. This framework has been introduced and comprehensively analyzed by Leone and collaborators in Ref.~\cite{leonetoappear}. 
The remaining operators, required to complete the total set of $|\Sigma_{k=3}(d=3)| = 8$ and $|\Sigma_{k=3}(d=5)|=12$ generators can be expressed  multiplicatively as  $A_\sigma \cdot Q_d $, with $A_\sigma$ the  representation of $\sigma \in \mathcal{S}_3$ in $\mathcal{H}_d^{\otimes k}$~\footnote{
These operators are not all independent, they can be rewritten in their irreducible representation, further simplifying computations. This reduces the dimension of the commutant—and consequently the size of the Gram and Weingarten matrices—from 8 to 7 for qutrits ($d=3$)  and 12 to 11 for qupents ($d=5$).}. 
We run simulations for several system sizes for qutrits ($d=3$) and qupents ($d=5$), computing the approach to the Haar value in Eq.~\eqref{eq:IPRHaarCl}. Specifically, we consider 
\begin{equation}
    \Delta \tilde{S}_3\equiv \log_d[ I^{\mathrm{Haar}}_3] -\log_d[\mathbb{E}_{\mathbf{C}_t}(I_3)],
    \label{eq:DeltaS3}
\end{equation}
and present this difference Fig.~\ref{fig:IPR}(a) and (b) for qutrits and qupents, respectively. 
Here, $\mathbb{E}_{\mathbf{C}_t}(I_k)$ is the average of the IPR evaluated at a certain time step $t$  evolving the circuit in Eq.~\eqref{fig:contracted_brickwall}. 
The exponential decay in time $\Delta \tilde{S}_3$, together with the uniform shift increasing system size $N$, are consistent with our expectation: higher order moments of the Clifford circuits saturate to their Haar value in depth logarithmic with the system size. 

As a complementary test, we consider now $d=2$ and study the overlap distribution of a shallow circuit with a small depth $t$, and compare it with Eq.~\eqref{eq:ziopera} with $x$ a fitting parameter. 
Our results for $N=128$ and $N=256$ two-qubit shallow circuits are presented in Fig.~\ref{fig:IPR}(c) and (d), respectively.
We see that, despite the limited circuit depth, the distribution Eq.~\eqref{eq:ziopera} captures quantitatively well the distribution of overlap of the Clifford circuit. 

These results substantiate our analytical conclusions. We finally conclude by commenting on the magic state resources required to recast the unitary Porter-Thomas distribution. 

\begin{figure*}[t!]
    \centering
    \includegraphics[width=\linewidth]{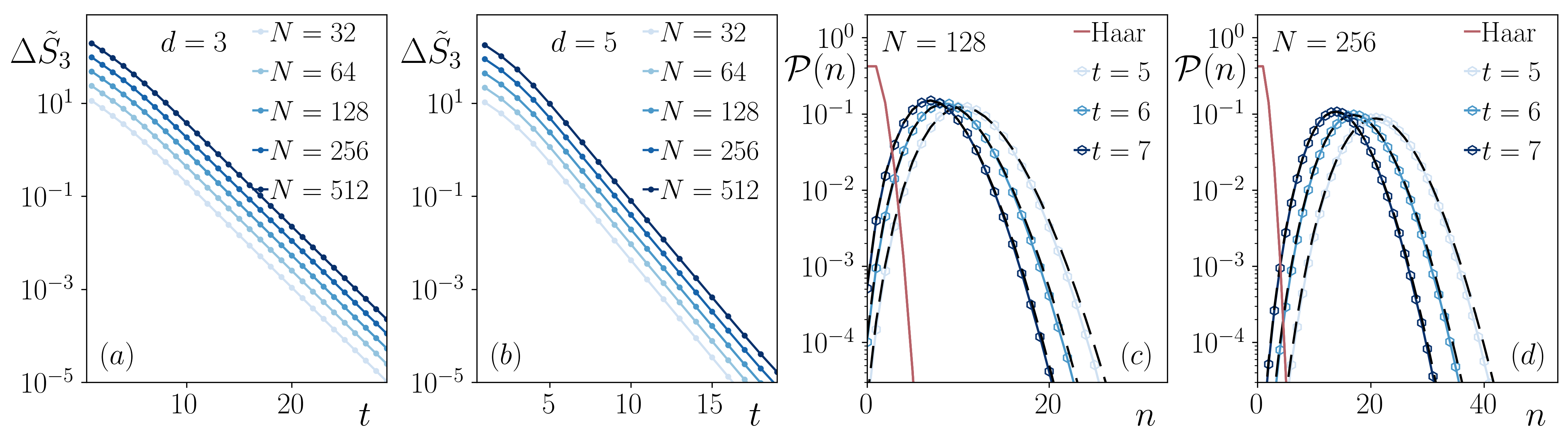}
    \caption{Dynamics of Eq.~\eqref{eq:DeltaS3} for qutrits $d=3$ (a) and qupents $d=5$ (b) for different system sizes. In both cases, the difference between the Haar value and the time-evolving IPR decreases exponentially fast with circuit depth, saturating for a fixed $\varepsilon$ in a timescale that is logarithmic in system size. 
    Distribution of the overlaps for shallow circuits with $t$ layers for $N=128$ (c) and $N=256$ (d), compared to Eq.~\eqref{eq:ziopera} where we fitted $x$ to match the most probable value of the distribution. In all cases, we consider  $\mathcal{N}=2\times 10^7$ data to sample the distribution, which is in quantitative agreement with our prediction. 
    }
    \label{fig:IPR}
\end{figure*}

\subsection{Beyond Clifford}
\label{sec:magic}
The discussion above demonstrates that Clifford circuits relax to the anticoncentration properties of random stabilizer states in logarithmic depth.
However, as noted earlier, Clifford transformations can be efficiently simulated on classical computers. The concept of non-stabilizerness, or “magic”, quantifies the non-Clifford resources required to achieve universal quantum computation. A state possesses magic resources if it does not belong to the set of stabilizer states.
There exist several measures of magic ($\mathcal{M}$) in many-body systems, often based on geometric distances~\cite{liu2022manybody} or the structure of the Clifford commutant \cite{Heinrich2019robustnessofmagic, lami2023nonstabilizerness, turkeshi2025pauli, leone2022stabilizer, piroli2021quantum}. 
In the following, the generalized stabilizer entropies $M_\sigma$, defined by \cite{turkeshi2025magicspreadingrandomquantum,leonetoappear}
\begin{equation}
    M_\sigma \equiv -\log[\zeta_\sigma],\quad\zeta_\sigma\equiv \left(\bigotimes_{i=1}^N\llangle \sigma_i|\right)|\Psi,\Psi\rrangle^{\otimes k}, 
\end{equation}
with $\sigma\in \overline{\Sigma}_k(d)\equiv \Sigma_k(d)\setminus \mathcal{S}_k$ non-permutation stochastic Lagrangian subspaces, will occur several times in our discussion. 
The quantity $\zeta_\sigma$ known as generalized stabilizer purity~\cite{leonetoappear}, corresponds to the expectation values of $ k $ copies of the state with elements of the Clifford commutant.
One can show that these quantities comply with magic measures requirements: 
(i) $M_\sigma(|\Psi\rangle)\ge0$ with equality holding if and only if $|\Psi\rangle$ is a stabilizer for $\sigma\in \overline{\Sigma}_k(d)$, (ii) they are invariant under Clifford unitary operations $M_\sigma(C|\Psi\rangle) = M_\sigma(|\Psi\rangle)$, and (iii) they are additive $M_\sigma(|\Psi\rangle\otimes |\Phi\rangle) =M_\sigma(|\Psi\rangle) +M_\sigma(|\Phi\rangle)$.

We now address a key question: how much non-Clifford resource is required for the system to recover the Porter-Thomas distribution of the Haar unitary ensemble? 
This problem is crucial, as anticoncentration to Haar statistics is a \textit{sine qua non} condition for achieving computational quantum advantage. 
To tackle this, we consider \textit{doped} Clifford circuits,  where a controlled amount of nonstabilizerness is injected in the system~\cite{Leone2021quantumchaosis,leone2024learning,Leone2024learningtdoped,lami2024learningstabilizergroupmatrix,haug2024probingquantumcomplexityuniversal,Chia2024efficientlearningof,bejain2024,szombathy2025independentstabilizerrenyientropy,szombathy2025spectralpropertiesversusmagic,niroula2023phase,Turkeshi2024,fux2024entanglement,tirrito2024quantifyingnon,hinsche2023one,zhou2020single,dowling2025bridgingentanglementmagicresources,fux2024disentanglingunitarydynamicsclassically, nakhl2024stabilizertensornetworksmagic}.
Concretely, we examine the behavior of random tensor networks and glued circuits acting on initial magic states. This setup is practically significant: since transversal non-Clifford gates are challenging to implement fault-tolerantly, magic state injection -- i.e., preparing a fault-tolerant logical qubit with magic—offers a more viable alternative for near-term quantum computing. 

\subsubsection{Doped random tensor networks}
We study the class of doped Clifford random matrix product states (dCRMPS), cf. Sub.\ref{subsec:resultsb}. 
These are obtained acting with the staircase circuit in Eq.~\eqref{eq:rmpsOBC}, on an initial state $|T_{(N_T)}\rangle = |T\rangle^{\otimes N_T}\otimes |0\rangle^{\otimes N-{N_T}}$, for a given magic state $|T\rangle$  such that ${M}_\sigma(|T\rangle) = \gamma_\sigma $ a constant $0<\gamma_\sigma\le 1$ for any $\sigma\in \overline{\Sigma}_k(d)$. It follows from the definition that $M_\sigma(|T_{(N_T)}\rangle) = N_T \gamma_\sigma$. 
We remark that, because of Clifford invariance, the nonstabilizerness of the initial state is preserved by Clifford evolution. However, as we demonstrate below, the anticoncentration properties will change substantially. 

Concretely, we answer how many $N_T$, in the jargon $T$-count, resources we need to recover the Porter-Thomas distribution. 
For this scope, let us first consider the solvable case represented in the quantum circuit representation by
\begin{equation}\label{eq:rmpsOBCTstate}
|\Psi^\mathrm{dCRMPS}_r\rangle\equiv\mathbf{C}|T_{(r)}\rangle=\begin{tikzpicture}[baseline=(current bounding box.center),scale=0.6,rotate=270]
   \definecolor{mycolor}{rgb}{0.6,0.8,0.9} 
    \foreach \x in {1,...,4}{
        \draw[thick, black] (\x,0) -- (\x,5.5);
        \draw[line width=0.4mm, fill=white] (\x,0) circle (0.2);
    }
    \draw[line width=0.4mm, fill=magenta] (1,0) circle (0.2);
    \foreach \x in {1,...,3}{
        \draw[line width=0.8mm, blue!60!black] (\x,1.5*\x-1.3) -- (\x,1.5*\x);
        \draw[line width=0.4mm, fill=mycolor, rounded corners=5pt] 
            (\x-0.2,1.5*\x-0.6) rectangle (\x+1.2,1.5*\x+0.2);
    }
    \draw[line width=0.8mm, blue!60!black] (4,1.5*4-1.3) -- (4,5.5);
\end{tikzpicture}\;,
\end{equation}
where $\mathbf{C}$ is the circuit in Eq.~\eqref{eq:rmpsOBC}, and the magenta dot represents the magic states $|T\rangle^{\otimes r}$ on the first virtual index. Note that in this setup, the parameter $r$ both controls the number $N_T = r$ of magic states and the bond dimension $\chi = d^r$.  
As a proxy of anticoncentration, we study the inverse participation ratios.

Upon contracting with the state $|\pmb{0}\rangle$ and performing the replica trick and the Clifford average, we obtain a circuit analogous to Eq.~\eqref{eq:rmpsOBC1}, which can be resummed exactly. The final result is given by 
\begin{equation}
\begin{split}
    I_k^\mathrm{dCRMPS}(r) &= d^{(1-k)N} \frac{\left[(-d^{-r};d)_{k-1}\right]^{N-r-1}}{\left[(-d^{-1-r};d)_{k-1}\right]^{N-r}}
    \left(\sum_{\sigma\in \Sigma_k(d)} \zeta_\sigma^{r}\right)
    \;.
\end{split}
\end{equation}
Furthermore, we can expand the second line using that for any state $|\phi\rangle$ in the Hilbert spaces, $\zeta_\sigma=1$ for any $\sigma\in \mathcal{S}_k$, so that $\sum_{\sigma\in \Sigma_k(d)} \zeta_\sigma^{r} = k! + \sum_{\sigma\in \overline{\Sigma}_k(d)} \zeta_\sigma^{r}$.  
The key observation is that generalized stabilizer purity $\zeta_\sigma\equiv d^{-M_\sigma}$ is a positive number between $1/d$ and $1$. 
First, let us consider the scaling limit $N,r\to\infty$ with $N/d^r=x$ constant. 
We recall that $I_k^\mathrm{Haar,\mathcal{U}}=k!/{\prod_{m=1}^{k-1}{(d^N+m)}}$, cf. App.~\ref{app:A} for a review of the anticoncentration of Haar unitary circuits. Then, a simple computation leads to 
\begin{widetext}
    \begin{equation}
    I_k^\mathrm{dCRMPS}(r) = I_k^\mathrm{Haar,\mathcal{U}} e^{\frac{d^k-d}{d^2}x} \left(1+\frac{\sum_{\sigma\in \overline{\Sigma}_k(d)}\zeta_\sigma^{\log_d(N/x_0)} }{k!}\right)
   =I_k^\mathrm{Haar,\mathcal{U}} e^{\frac{d^k-d}{d^2}x} \left[1+\sum_{\sigma\in\overline{\Sigma}_{k}(d)}\left(\frac{x_0}{N}\right)^{M_\sigma}\right]\mapsto_{N\to\infty} I_k^\mathrm{Haar,\mathcal{U}} e^{\frac{d^k-d}{d^2}x},
   \label{eq:rmpsIPRdoped}
\end{equation}
\end{widetext} 
where $x$ is given by Eq.~\eqref{eq:ziioooo} with $x_0 = N/\chi$ constant.
In essence, this result demonstrates how doped Clifford random matrix product states approximate the Porter-Thomas distribution up to Poisson fluctuations. 
We test this result in Fig.~\ref{fig:dopednumerics}(a), focusing on $k=3$ and $d=3$~\footnote{Similar tests and results were corroborated for other choices of $k$ and $d$, and are not reported for presentation purposes.} and computing 
\begin{equation}
    \Delta \tilde{S}_3^\mathrm{U}\equiv \log_d[ I^{\mathrm{Haar},\mathcal{U}}_3] -\log_d[I_3^\mathrm{dCRMPS}(r)].
    \label{eq:DeltaS32}
\end{equation}
Our results present substantial agreement with the Poisson prediction. For reference, we compare this result with the log-normal corrections characteristic of chaotic systems~\cite{lami2025anticoncentration,christopoulos2024universal}, which are in agreement with the data only at $x\sim 0$.
Since $x_0= N/\chi$, this limit corresponds to injected magic resources growing as $O(\log(N))$. 

\begin{figure*}[t!]
    \centering
    \includegraphics[width=0.7\linewidth]{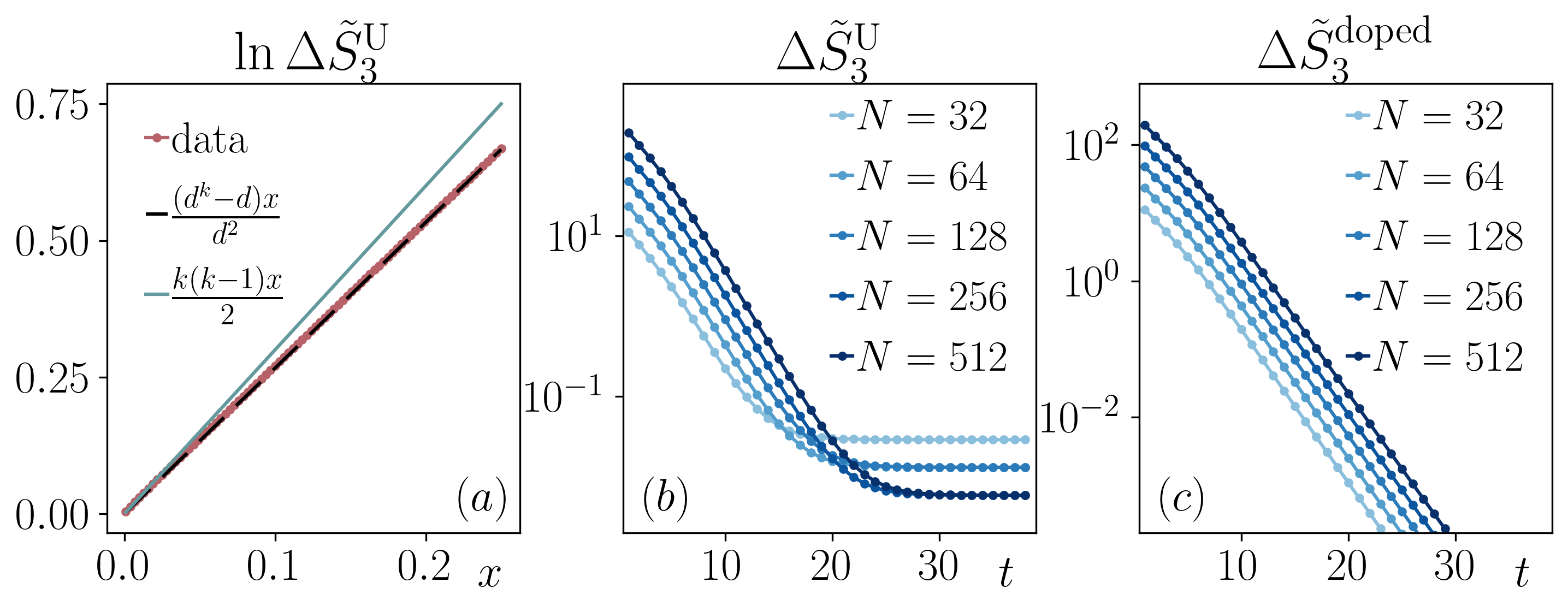}
    \caption{(a) Fluctuations between the doped random matrix product state $I_3$ and the Haar unitary value for different $N\le 512$ and $4\le r\le 12$ compared to the log-normal fluctuations expected for unitary circuits, and the Poisson fluctuations of Clifford circuits. The data displays strong agreement with the latter, corroborating Eq.~\eqref{eq:rmpsIPRdoped}. 
    (b) Evolution of Eq.~\eqref{eq:DeltaS32} under random circuits of qutrits ($d=3$) for different system sizes starting from a system with a number $N_T=\lfloor \log_2(N)/2\rfloor$ of $|T\rangle$ states. This quantity decreases exponentially fast in system size up to a convergence in $\log(N)$ depth toward a value that $\Delta\tilde{S}^\mathcal{U}_3(\infty)$ that decreases polynomially in system size. 
    (c) Evolution of Eq.~\eqref{eq:maria} with respect to the doped global Clifford value Eq.~\eqref{eq:maria2}. This quantity decreases exponentially fast in system size, saturating in a timescale that is logarithmic in system size $\log(N)$. 
    }
    \label{fig:dopednumerics}
\end{figure*}

\subsubsection{Doped Clifford circuits}
We now extend our considerations to doped Clifford circuits. 
First, we consider the doped glued circuit (dGC), obtained acting the Clifford glued circuit in Eq.~\eqref{fig:stitched_circuit} on the initial state $|T_{(2r)}\rangle$ on the first gate. 
\textit{Mutatis mutandis}, we can again resum the expression exactly, obtaining 
\begin{equation}
\begin{split}
    I_k^\mathrm{dGC}(r) &= d^{(1-k)N} \frac{\left[(-d^{-r};d)_{k-1}\right]^{\frac{N-r}{r}}}{\left[(-d^{-2r};d)_{k-1}\right]^{\frac{N-2r}{r}}}\\ &\quad \times \left[k!+\sum_{\sigma\in \overline{\Sigma}_k(d)} \zeta_\sigma^{2r}
    \right]\;.
\end{split}
\end{equation}
Taking the scaling limit with $x_0 = \frac{d}{d-1} \frac{N}{r d^r}$ and $N,r\to\infty$, we have 
\begin{equation}
    I_k^\mathrm{dGC}(r) \mapsto_{N\to\infty} I_k^\mathrm{Haar,\mathcal{U}} e^{\frac{d^k-d}{d^2}x}\;,
    \label{eq:dopedstitched}
\end{equation}
with $x = x_0(1-2r/N)+O(x_0^2)$.
As a result, shallow circuits at depth $O(\log(N))$ with a number of $N_T=O(\log(N))$ magic state resources retrieve the Haar unitary anticoncentration properties.

The above discussion holds for the thermodynamic limit. In a practical setup, it is important to consider finite-size systems and actual implementation.   
For this task, it is crucial the concept of a physical observer, an agent that can resolve the difference between two quantities with at most polynomial resources in system sizes.
Our analysis for doped Clifford tensor networks and circuits implies that any $r=O(\log^{1+c}(N))$ for $c>0$ suppress faster than any polynomial the difference between the Haar anticoncentration value. 
These results suggest that doped Clifford tensor network and doped Clifford shallow circuits are pseudo-magic quantum states \cite{gu2024pseudomagic}, analogous to shallow circuits being pseudo-entangled states~\cite{schuster2025randomunitariesextremelylow,liu2024quantum, cheng2024pseudoentanglement,feng2024dynamic}.
However, it is important to remark that achieving pseudo-randomness generally requires circuit depths of at least $\Omega(N)$ magic resources~\cite{schloss,haug2024pseudorandomunitariesrealsparse}. 

We conclude this section, by benchmarking the core result Eq.~\eqref{eq:rmpsIPRdoped} with Clifford tensor replica network methods for a local circuit similar to Eq.~\eqref{fig:brickwall}. 
We focus on $d=3$ and $k=3$~\footnote{We have checked but not reported that similar results hold for $k=3$ and $d=5$ qudits, and for qubits ($d=2$) and $k=4$.}, initializing the state in $|T_{(N_T)}\rangle$ with $|T\rangle =( |0\rangle + e^{2\pi i/9} |1\rangle + e^{-2\pi i/9} |2\rangle)/\sqrt{3}$. 
The $T$-count is fixed by $N_T = \lfloor\log_2(N)/2\rfloor$ with $\lfloor x\rfloor$ the integer truncation of $x$.
As a measure of anticoncentration, we consider Eq.~\eqref{eq:DeltaS32} and present the results obtained via replica tensor network for $32\le N\le 512$ in 
Fig.~\ref{fig:dopednumerics}(b). 
We see that the distance between the participation entropy decreases exponentially fast in circuit depth $t$, upon saturating to a value $\Delta \tilde{S}_3(\infty)=N^{-c}$ that decreases polynomially in system size. 
This mismatch arises from finite-size corrections. The stationary state reached at late times can be approximated by replacing the full evolution with a random global Clifford, allowing for an analytical determination of the resulting value
\begin{equation}
    I_3^\mathrm{doped,\mathrm{Haar}} \simeq d^{(1-k) N} \left[6+ 2\left(\frac{2}{3}\right)^{N_T}\right]\;,\label{eq:maria2}
\end{equation}
where we neglected exponentially small corrections. 
In Fig.~\ref{fig:dopednumerics}(c) we test this conclusion is correct by computing
\begin{equation}
    \Delta \tilde{S}_3^\mathrm{doped}\equiv \log_d[ I^{\mathrm{doped},\mathrm{Haar}}_3] -\log_d[\mathbb{E}_{\mathbf{C}_t}(I^\mathrm{doped}_3)]\;,\label{eq:maria}
\end{equation}
which is exponentially decaying with $t$ toward zero. 
These results provide further validation of our analytical predictions.

\section{Discussion}
\label{sec:disc}

This work examined the anticoncentration properties of Clifford circuits and the role of non-Clifford resources in restoring the Porter-Thomas distribution of unitary random states. 
After identifying the overlap distribution of random stabilizer states, we demonstrated that Clifford tensor network states reach this distribution with a bond dimension growing linearly with system size. 
Building on these results, we established a mapping to glued circuits, revealing that shallow circuits of depth $O(\log(N))$ anticoncentrated to random stabilizer states. Crucially, this timescale is much faster than the condition to reach approximate Clifford designs~\cite{haferkamp2022}.
Furthermore, we demonstrated that recovering the overlap distribution characteristic of quantum chaotic states required a $T$-count that scaled logarithmically with system size. 
To support our findings, we performed extensive numerical simulations using tableau methods and introduced the Clifford replica tensor network, providing strong evidence for our analytical predictions.

The collection of techniques, frameworks, and results developed in this work presents several relevant implications in quantum information science and beyond. 
We briefly comment on some of these directions and present a list of prominent outlooks. 

\subsection{Implications for quantum state design}
The anticoncentration properties of shallow Clifford circuits extend naturally to quantum state design, as quantified by the frame potential~\cite{christopoulos2024universal,lami2025anticoncentration}. This quantity measures how closely a given ensemble of quantum states approximates the uniform Haar distribution over the Hilbert space. The $k$-th frame potential is defined as
\begin{equation}
    F_k = \mathbb{E}_{\psi, \psi'}\left[ |\langle \psi' | \psi \rangle |^{2k} \right],
\end{equation}
where the expectation is taken over pairs of independent states sampled from the same ensemble. 
Consider states generated by shallow Clifford circuits, $|\psi\rangle = U_t|0\rangle$ and $|\psi'\rangle = U_t'|0\rangle$, where $U_t$ and $U_t'$ are independent random circuits of depth $t$. 
Setting $V_{2t} = (U_t')^\dagger U_t$, the frame potential can be rewritten as the ensemble average of the overlap $F_k = \mathbb{E}_{V_{2t}} \left[ |\langle 0 | V_{2t} | 0 \rangle |^{2k} \right] \propto I_k$,
where $V_{2t}$ is now a circuit of depth $2t$. Hence, if Clifford circuits anticoncentrate at logarithmic depth, the frame potential approaches its (Clifford) Haar value $F_k \to F_k^{\mathrm{Haar}}$ in depth $O(\log N)$. 
This implies that (additive) stabilizer state designs can be constructed efficiently using logarithmic-depth circuits.

Furthermore, injecting a polylogarithmic amount of magic into the initial state enables the ensemble to approximate a Haar-random circuit ensemble, yielding
\begin{equation}
    F_k^{\mathrm{doped}} = \mathbb{E}_{V_{2t}} \left[ |\langle T_{(r)} | V_{2t} | T_{(r)} \rangle |^{2k} \right] \to F_k^{\mathrm{Haar}, \mathcal{U}}.
\end{equation}
where $|T_{(r)}\rangle$ denotes the magic input state in Sec.~\ref{sec:magic}. 
Achieving a relative state design, however, remains more challenging. 
Recent work~\cite{grevink2025glueshortdepthdesignsunitary} demonstrated that the gluing lemma fails for stabilizer ensembles: for qubits, a circuit of linear depth is required to achieve a relative state design for any $k \geq 3$. 
This leaves open the question of how many magic-state resources are necessary to reach relative designs approximating Haar-random unitary states. 
These considerations have direct implications for shadow tomography in prime qudit systems with dimension $d \geq 3$, as we detail in a forthcoming publication~\cite{magni2025}.


\subsection{Implications in quantum many-body physics} 
Our work extends, in the context of Clifford circuits, recent analyses on the emergence of universal behavior in finite-depth quantum circuits~\cite{deluca2024universalityclassespurificationnonunitary,lami2025anticoncentration,nahum2017quantum}. 
As in those studies, averaging over circuit ensembles leads to a natural mapping onto effective statistical physics models, where the local degrees of freedom correspond to elements of the commutant algebra. 
In this framework, universality emerges in regimes where domain walls—or “kinks”—between different commutant sectors are sparsely distributed~\cite{deluca2024universalityclassespurificationnonunitary}.  
In the case of Haar-random unitary circuits, similar techniques have been applied to study the spectral properties of Floquet dynamics~\cite{chan2018,Chan2022}, revealing a universal crossover to quantum chaos governed by random matrix theory. 
Our results highlight that Clifford circuits exhibit a distinct form of random matrix universality in their overlap distributions—one that lies beyond the standard Dyson ensembles of unitary, orthogonal, or symplectic types~\cite{christopoulos2024universal,Braccia2024computing,sauliere2025universalityanticoncentrationchaoticquantum}.

From this perspective, the spectral analysis of Clifford circuits presents a compelling opportunity: can standard indicators of quantum chaos—such as level repulsion—offer insight into the classical simulability of Clifford circuits? 
Moreover, does the crossover to the Porter-Thomas distribution, controlled by the degree of $T$-gate doping, have a counterpart in the spectral correlations? 
Our results, together with the theoretical framework developed in this work, open the door to addressing such questions at the intersection of practical quantum computation, classical and quantum complexity, and the physics of quantum chaos. 
A promising route to explore these directions is a perturbative analysis of Clifford domain walls. 
This could help resolve several open problems—for example, understanding why the universal spectral fluctuations typically expected in Haar-random unitary circuits appear compatible with Clifford circuit numerics, as reported in Ref.~\cite{sierant2023ent}.


\subsection{Implications in quantum circuit sampling and benchmarking of NISQ}
A more practical implication concerns the benchmarking of quantum computational advantage, a central challenge in the NISQ era. 
Here, the goal is to demonstrate that a quantum device can perform a computational task beyond the capabilities of any classical computer. 
A key example is random circuit sampling, where one tests whether the device generates output distributions that exhibit signatures of quantum complexity, such as the Porter–Thomas (PT) distribution. 
The fidelity of such output can be assessed via cross-entropy benchmarking (XEB), comparing the experimental data to classical simulations. 
Reaching the PT regime is thus a practical proxy for establishing computational hardness and validating device performance.

As discussed, in near-term platforms, doped Clifford framework—wherein Clifford circuits are sparsely augmented with non-Clifford gates—offers a flexible and physically motivated setting to explore this complexity transition. By tuning the density and placement of magic gates, one can interpolate between classically tractable and computationally hard regimes. 
Our results show that doped Clifford circuits reach the regime of approximate unitary Porter-Thomas statistics with circuit depth and number of magic gates scaling only polylogarithmically with system size. 
This scaling enables the simulation and benchmarking of large-scale quantum circuits using efficient tableau-based stabilizer techniques, which are well beyond the reach of tensor network methods or full universal gate sets. 
Importantly, this provides a controlled and physically motivated setting to validate quantum complexity in regimes that are provably classically hard, offering a scalable diagnostic framework for near-term quantum advantage experiments.

\subsection{Outlook}
Overall, our results highlight the interplay between Clifford dynamics, quantum complexity, and randomness generation, providing new insights into the structure of quantum circuits and their computational power. 
These findings open the door to several promising research directions. One key avenue is exploring how different geometries affect anticoncentration. As in the unitary Haar case~\cite{christopoulos2024universal}, we expect such variations to alter the nature of leading fluctuations in the associated statistical mechanics model. 
More broadly, investigating the effects of noise in doped Clifford circuits is crucial for connecting our results to noisy intermediate-scale quantum (NISQ) devices. Similar analyses in Haar unitary circuits~\cite{fefferman2024anticoncentrationunitaryhaarmeasure,fefferman2024effect,jian2022linear} suggest that noise could play a significant role in shaping anticoncentration properties.

\begin{acknowledgments}
We thank L. Leone, and S. Pappalardi for insightful discussions, and G. Lami, J. De Nardis and E. Tirrito for collaborations on related topics. We are particularly grateful to N. Dowling, M. Heinrich and P. Sierant for discussion and comments on the manuscript. 

B.M. and X.T. acknowledge DFG Collaborative Research Center (CRC) 183 Project No. 277101999 - project B01 and DFG Germany's Excellence Strategy – Cluster of Excellence Matter and Light for Quantum Computing (ML4Q) EXC 2004/1 – 390534769. ADL acknowledges support by the
ANR JCJC grant ANR-21-CE47-0003 (TamEnt). A.C acknowledges support by the  ERC StG 2022 project DrumS, Grant Agreement 101077265.

\textbf{Data availability.}
The numerical data for this work are given in Ref.~\cite{data}, available at publication.

\end{acknowledgments}

\appendix 
\section{Anticoncentration of Haar unitary circuits} \label{app:A}

We summarize key results on anticoncentration in Haar-random unitary circuits, which provide a baseline for comparison with our findings on Clifford circuits and doped Clifford circuits. 

A random Haar state in the Hilbert space is obtained acting with a Haar-random unitary matrix $U$ acting on an initial computational basis state $|\pmb{0}\rangle $, namely  $|\Psi\rangle = U |\pmb{0}\rangle$. 
As anticipated in the Main Text, the distribution of overlaps $w = d^N|\bra{\Psi}\pmb{0}\rangle|^2 $  for this ensemble of states follows the Porter-Thomas distribution~\cite{porter1956fluctuations}
\begin{equation}
P(w) = \frac{d^N-1}{d^N}\left(1-\frac{w}{d^N}\right)^{d^N-2}\mapsto_{N\gg1} e^{-w}.
\end{equation}
From this distribution, we can compute the inverse participation ratios by  
\begin{equation}
I_k^\mathrm{Haar,\mathcal{U}} = d^{(1-k)N}\mathbb{E}[w^k] = \frac{k!}{\prod_{m=0}^{k-1}{(d^N+m)}},
\end{equation}
where $k!$ is the dimension of permutation group $\mathcal{S}_k$. 
In the large $N$ limit, this simplifies to  
\begin{equation}
    I_k^\mathrm{Haar,\mathcal{U}} \approx k! d^{(1-k)N}.
    \label{eq:haarval}
\end{equation}

In chaotic quantum circuits, locality plays a crucial role, requiring a threshold depth to reach the Porter-Thomas distribution. Random unitary matrix product states converge to this distribution with a logarithmic bond dimension in the number of qudits~\cite{lami2025anticoncentration}, while quantum circuits saturate to the Haar unitary value at a depth logarithmic in the number of qudits~\cite{christopoulos2024universal}. 
In both cases, the Haar unitary inverse participation ratios $I_k^{\mathcal{U}}$ can be expressed as the product of the Haar value~\eqref{eq:haarval} and a log-normal correction
\begin{equation}
I_k^{\mathcal{U}} = I_k^{\mathrm{Haar},\mathcal{U}} e^{\frac{k(k-1)}{2}x}\;.
\end{equation}
The explicit overlap distribution for generic circuits was also derived in~\cite{lami2025anticoncentration,christopoulos2024universal}, provided $x$ can play the role of a fitting parameter.

\bibliography{filtered_bib}

\end{document}